\documentclass[prb,aps,twocolumn,showpacs,superscriptaddress]{revtex4}

\usepackage{graphicx}
\usepackage{bm}
\usepackage{amssymb}
\usepackage{amsmath}
\usepackage{color}
\usepackage{ulem}

\def \be {\begin{equation}}
\def \ee {\end{equation}}
\def \bea {\begin{align}}
\def \eea {\end{align}}
\def \p {\partial}
\def \BEA {\begin{eqnarray}}
\def \EEA {\end{eqnarray}}

\def \BC {\begin{cases}}
\def \EC {\end{cases}}
\def \p{\partial}
\newcommand{\beml}{\begin{subequations}}
\newcommand{\eml}{\end{subequations}}
\newcommand{\beq}{\begin{eqnarray}}
\newcommand{\eq}{\end{eqnarray}}

\begin{document}

\title{Resonant inverse Faraday effect in nanorings}

\author{K.\,L.~Koshelev}
\affiliation{A.\,F.~Ioffe Physico-Technical Institute, 194021 St.~Petersburg, Russia}
\affiliation{Radboud  University,  Institute  for  Molecules  and  Materials,  NL-6525  AJ  Nijmegen,  The  Netherlands}
\affiliation{ITMO University, 197101 St.~Petersburg, Russia}
\affiliation{Peter the Great St. Petersburg Polytechnic University, 195251 St.~Petersburg, Russia}
\author{V.\,Yu.~Kachorovskii}
\affiliation{A.\,F.~Ioffe Physico-Technical Institute, 194021 St.~Petersburg, Russia}
\affiliation{Radboud  University,  Institute  for  Molecules  and  Materials,  NL-6525  AJ  Nijmegen,  The  Netherlands}
\author{M.~Titov}
\affiliation{Radboud  University,  Institute  for  Molecules  and  Materials,  NL-6525  AJ  Nijmegen,  The  Netherlands}

\begin{abstract}
A circularly polarized light can induce a dissipationless dc current in a quantum nanoring which is responsible for a resonant helicity-driven contribution to magnetic moment.  This current is not suppressed by thermal averaging despite its quantum nature. We refer to this phenomenon as the quantum resonant inverse Faraday effect. For weak electromagnetic field, when the characteristic coupling energy is small compared to the energy level spacing, we predict narrow resonances in the circulating current and, consequently, in the magnetic moment of the ring. For strong fields, the resonances merge into a wide peak with a width determined by the spectral curvature.  We further demonstrate that weak short-range disorder splits the resonances and induces additional particularly sharp and high resonant peaks in dc current and magnetization. In contrast, long-range disorder leads to a chaotic behavior of the system in the vicinity of the separatrix that divides the phase space of the system into regions with dynamically localized and delocalized states.

\end{abstract}

\pacs{78.20.Ls, 78.67.-n, 73.23.-b, 75.75.-c}

\maketitle

\section{Introduction}

Nanodevices based on quantum dots, quantum wires and quantum rings continue to attract considerable attention.\cite{Dai15,Liu15,Kostarelos14,Jayich09,Birge09,Srivastava15,Fang15,Beaulac09,Cave09} From the physics point of view such systems are often determined by the interplay of quantum interference and charge quantization effects which both become more prominent with decreasing system size and temperature. Research on electronic phenomena such as the Aharonov-Bohm effect, Anderson localization, Kondo effect, or Coulomb blockade has been dominating the field in the last two decades.\cite{Borunda08,Oudenaarden98,Webb85,Schoenenberger99,Grbic08,Titov97,Aleiner02,Evers08,Kouwenhoven98} In recent years, however, there have appeared numerous proposals to utilize nanodevices in optoelectronics and spintronics.\cite{Srivastava15,Fang15,Beaulac09,Cave09,Borunda08}
This development calls for better understanding of light-matter interaction in such systems as quantum wire antennas, artificial atoms, and nanorings.\cite{Oudenaarden98, Webb85, Schoenenberger99, Grbic08, Titov97}

One of the main goals of optoelectronics is to design and fabricate tunable electronic nanodevices that are capable of operating in a frequency range unaccessible for conventional electronic technologies, i.e. in the so-called terahertz (THz) gap. It is widely believed that the frequency gap can be closed using optoelectronic and plasmonic devices. There is, however, a serious obstacle for such development. The coupling of THz electromagnetic field to a single nanosystem appears to be too weak because the typical dimension of a nanosystem is two or more orders of magnitude smaller than the THz wavelength. A promising way to increase the coupling is to use periodic structures (arrays of nanoparticles, grating gate structures, multigate structures, etc). Another difficulty originates in dc photoresponse that is only possible in the presence of a system asymmetry (which would define the direction of the photoinduced dc current). In two-dimensional  systems such an asymmetry might be created by boundary conditions \cite{Dyakonov93} or induced by a ratchet effect (see Ref.~\onlinecite{Ivchenko2011} for review) . The latter implies a special type of grating-gate couplers that could provide the required asymmetry.

Interestingly, the symmetry conditions for photoresponse are more relaxed in the multiconnected structures such as quantum rings. In particular, the dc circular current can be excited in a quantum ring by a circularly polarized optical field: $\bm{\mathcal{E}}= \bm{\mathcal{E}}_\omega \exp(-i\omega t )+ h.c.$, where $\bm{\mathcal{E}}$ is the electric-field component of the electromagnetic wave. Such response can be characterized by an orbital magnetic moment of the ring
\be
\boldsymbol{M}  \propto    i\; \bm{\mathcal{E }}_\omega \times \bm{\mathcal{ E}}^*_\omega,
\label{M}
\ee
the effect which is commonly referred to as the inverse Faraday effect.\cite{Pitaevskii61,Ziel65,Kimel05} In contrast to other photomagnetic effects, the inverse Faraday effect does not involve absorption of photons or heating, which makes it particularly useful for spintronic applications such as data storage technologies.\cite{Kirilyuk11}  Although the magnetic moment generated in a single ring is relatively small, an ensemble of nearly identical quantum rings may give rise to large optically controlled macroscopic magnetization. Two points are especially important in view of possible applications: (i) the  proportionality coefficient in Eq.~\eqref{M}  is an odd function of frequency, so that the effect is sensitive to the helicity of polarization, and (ii) the effect is sizable even in the limit of long wavelength such that $\bm{\mathcal{E }}_\omega$  does not vary within the ring dimension. Hence, quantum nanorings and ring-based arrays can be used as effective helicity-driven sensors for THz radiation.

Historically, the inverse Faraday effect has been predicted by Pitaevskii\cite{Pitaevskii61} and first observed by van der Ziel \textit{et al.}\cite{Ziel65} Much of the current interest to the phenomenon originates, however, in the experiments by Kimel \textit{et al.}\cite{Kimel05,Kirilyuk10,Kirilyuk11} on ultrafast  femtosecond magnetization dynamics in thin ferrimagnets.  In this paper we leave aside many unresolved issues in the theory of the inverse Faraday effect in magnetic materials but focus instead on the excitation of current and magnetic moment in a single-channel quantum ring. This problem has been analyzed recently by Kibis \cite{Kibis11} using perturbative analysis (see also a more recent publication \cite{Kibis13}) and by Alexeev \textit{et al.}\cite{Alexeev13} using a master equation while disregarding diamagnetic current.    A similar system but in the presence of strong spin-orbit interaction at zero temperature has been recently considered.\cite{Joibari14}  The electric dipole moment oscillations in quantum rings at a finite temperature \cite{Alexeev12} and inverse Faraday effect due to the flux change through classical (macroscopic) metallic rings subjected to short optical pulses were also discussed.\cite{Kruglyak2005,Kruglyak2007}  The inverse Faraday effect in mesoscopic chaotic cavities has been studied in Ref.~\onlinecite{Polianski2009}. In this paper we focus on the inverse Faraday effect in quantum rings. In contrast to previous publications we develop a nonperturbative approach that remains valid for the case of strong coupling to electromagnetic field.  We focus specifically on the resonant enhancement of the inverse Faraday effect in the absence of spin-orbit interaction and at relatively high temperatures.

Optically-induced circular current $I_{\rm rad}$ has a number of similarities to persistent current $I_{\rm per }$. The latter may flow in a quantum ring at thermodynamic equilibrium.  Both currents are dissipationless  and vary periodically with magnetic flux piercing the ring. Both currents arise due to time-reversal symmetry breaking by magnetic field and/or by circularly polarized light. Consequently,  $I_{\rm per }$ is an odd function of the magnetic field $I_{\rm per}(\phi)=-I_{\rm per} (-\phi)$, while $I_{\rm rad} $ changes sign upon the inversion of both magnetic field and optical field helicity
\be
I_{\rm rad}(\phi,\omega)=-I_{\rm rad} (-\phi,-\omega) .
\label{Irad_phi_omega}
\ee
Here we introduce $\phi=\Phi/\Phi_0$, where $\Phi$ is the magnetic flux piercing the ring and $\Phi_0=hc/e$ is the flux quantum.

The persistent and optically induced currents are, however, completely different when it concerns their temperature dependence. The averaged persistent current is exponentially suppressed with increasing temperature $T$ above the level spacing at the Fermi level $\Delta_F$,\cite{Imry}  namely $I_{\rm per}\propto \exp(-T/\Delta_F)$, while the optically induced current decays much slower, $I_{\rm rad}\propto \Delta_F/T$, as we show below.  Another closely related difference is related to the role of mesoscopic fluctuations in these two currents. Such fluctuations provide a dominant contribution to the persistent current for all temperatures, which makes it sensitive to the type of the thermodynamic statistical ensemble.\cite{Altshuler91} (Since fluctuations exponentially exceed the averaged value of the current, the quantity $I_{\rm per}$  is not representative for a given isolated ring.)

In contrast, as we demonstrate below, the mesoscopic fluctuations of $I_{\rm rad}$ are small for temperatures exceeding the mean level spacing so that the ensemble-averaged optical current is well defined. The dependence of $I_{\rm rad}$ on the type of thermodynamic averaging is, therefore, negligible at high temperatures.

To conclude the  comparison  of  $I_{\rm per}$ and $I_{\rm rad}$,  we note that persistent current might show up indirectly at high temperatures.  In particular, it was demonstrated in a series of publications \cite{Dmitriev10,Shmakov12,Dmitriev14} that the tunneling current through a single-channel quantum ring is blocked by persistent current. This effect, caused by an interplay of quantum interference and charge quantization, has been named the persistent-current blockade (PCB) in analogy with the well-known Coulomb blockade. In contrast to the latter, the PCB persists for much higher temperatures despite its essentially quantum nature. The mesoscopic fluctuations of $I_{\rm per}$ in the regime of PCB lead to the splitting of Aharonov-Bohm resonances at high temperatures. Similarly, we will find that the current $I_{\rm rad}$ survives up to sufficiently large temperatures.

The slow decay of the inverse Faraday effect with temperature yields additional advantages for optoelectronics and spintronics.  Therefore, in this paper we focus on the high-temperature regime
\be
T\gg\Delta_F.
\label{TDF}
\ee
We calculate $I_{\rm rad}$ for arbitrary coupling to electromagnetic radiation by paying a particular attention to resonance effects. In the weak field limit we predict series of narrow resonances in the frequency dependence of $I_{\rm rad}$. Each resonance corresponds to excitation frequency coinciding with the distance between neighboring levels. Weak short-range disorder splits the resonances and induces particularly sharp resonant peaks in magnetization. For the case of large field we find using the quasiclassical approximation that the resonances broaden and merge into a single wide peak. The width of the peak is limited by a thermal band for moderate coupling while it is proportional to the square root of the wave amplitude for very strong fields. In a clean limit, i.e., in the absence of disorder, the corresponding circular current is dissipationless. The presence of long-range disorder leads to a chaotic behavior of the system in the vicinity of the separatrix that divides the phase space into regions with dynamically localized and delocalized states.  In contrast, weak short range disorder leads to the appearance of additional particularly sharp and high resonant peaks in dc current and magnetization.

\section{Ideal quantum ring}\label{single}

An ideal quantum ring placed in a circularly polarized electromagnetic field can be described with an effective stationary Schr{\"o}dinger equation by transforming to the rotating frame. The solution to this equation is straightforwardly obtained in two limiting cases: (i) for weak electromagnetic field, such that the coupling energy is small compared to the level spacing, and (ii) for strong field, such that the level spacing in the ring is negligible compared to the coupling matrix element between electrons and photons.

Before going into details let us recall first the well-known concept of the persistent current in a ballistic single-channel ring.  In the absence of both electromagnetic radiation and magnetic flux each energy level in an ideal ring carries the electric current $I_n=I_0 n$, where
\be
I_0=\frac{e\hbar}{2\pi MR^2}.
\ee
 Here $R$ stands for the radius of the ring, $M$ is an effective electron mass, and integer number $n$ numerates energy levels. Due to the evident symmetry $I_n=I_{-n}$  the total equilibrium current circulating in the ring vanishes. If the ring is threaded by a magnetic flux one finds
\be
I_n=I_0 (n-\phi). \label{In}
\ee
Thus, for generic flux the exact cancellation is absent and a dissipationless persistent current flows. At zero temperature one estimates the persistent current as
\begin{equation}
\label{pers}
I_\textrm{per}=\sum_{E_n^{(0)}<E_F}\!\!\!\!\! I_n \;\neq 0,
\end{equation}
where $E_n^{(0)}$ are energy levels in the absence of radiation and $E_F$ is the Fermi energy.  It is evident from Eq.~\eqref{pers} that the persistent current flows even in the absence of an external electric or electromagnetic field.

The persistent current (\ref{pers}) has been indeed observed in experiments with ensembles of nanorings.
\cite{Levy90,Reulet95,Deblock02,Kleemans07} (see also Ref.~\onlinecite{Imry} for review).
 Quantitative theoretical explanation of these experiments is, however, much more involved given that the rings are typically disordered and not one-dimensional while the effects of electron-electron interactions are not negligible.

Nevertheless, we shall start with the discussion of the simplest model, which is a clean single-channel quantum ring, and postpone the generalization of our results for the disordered case to Sec.~\ref{disorder}. The case of a multichannel ring as well as the effect of electron-electron interaction will be discussed qualitatively in  Secs.~\ref{many} and \ref{discussion}.

%%%%%%%%%%%%%%%
% Figure 1, fig:ring
%%%%%%%%%%%%%%%
\begin{figure}[t]
\centerline{\includegraphics[width=0.3\textwidth]{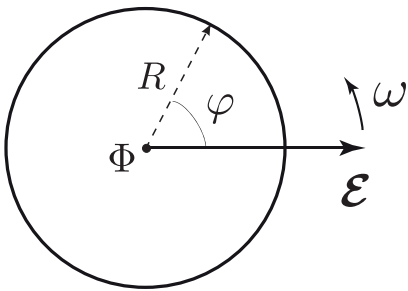}}
\caption{Single-channel quantum ring threaded by magnetic flux $\Phi$. Circularly polarized light induces a nondissipative current circulating around the ring.}
\label{fig:ring}
\end{figure}
%%%%%%%%%%%%%%%

We consider a single-channel nanoring subject to a circularly polarized radiation with frequency $\omega$. The radius of the ring is naturally assumed to be small  compared to the wavelength of light. In this case the electric field acting on the electrons in the ring is homogeneous and is given by
\be
\bm{\mathcal E}_\omega \approx {\mathcal E_0} {(\mathbf e_x- i \mathbf e_y) }/2,
\ee
where $\mathbf e_x$ and  $\mathbf e_y$ are unit vectors in $x$ and $y$ directions, respectively, and $\mathcal{E}_0$ is the amplitude of the field. The Schr{\"o}dinger  equation for the ring, which is threaded by a magnetic flux, is given by
\begin{equation}
\label{Koshelev1}
i\hbar\frac{\partial\Psi}{\partial t}=-\frac{\varepsilon_0}{2}\left(\frac{\partial  }{\partial \varphi }-i\phi\right)^2\Psi-e{\cal E}_0R\cos{(\varphi-\omega t)}\Psi,
\end{equation}
where $\varepsilon_0= \hbar^2/MR^2$ and $\varphi$ is the polar angle shown in Fig.~\ref{fig:ring}. In Eq.~\eqref{Koshelev1} we neglect small corrections arising due to a finite size of electron wave function in the radial direction. The function $\Psi$ corresponds to a state which carriers a dc current given by
\be
I=2\pi I_0 \left \langle  \left[\frac{1}{2i}\left(\Psi^*\frac{\partial \Psi}{\partial  \varphi} -\Psi \frac{\partial \Psi^*}{\partial  \varphi}\right) -\phi|\Psi|^2\right] \right\rangle_t,
\label{current00}
\ee
where $\langle \cdots \rangle_t$ stands for time averaging.

Equation \eqref{Koshelev1} can be transformed into a stationary Schr{\"o}dinger equation using a rotating reference frame,
\begin{equation}
\label{Koshelev2}
\Psi(\varphi,t)=e^{-iEt} e^{i\phi\varphi} e^{i n_\omega(\varphi-\omega t)}\chi{(\varphi-\omega t)},
\end{equation}
where we introduce the dimensionless frequency and coupling
\be
n_\omega=\frac{\omega}{\varepsilon_0},\quad
\alpha=\frac{ e{\cal E}_0R}{\varepsilon_0}.
\label{Koshelev3}
\ee
Here and in what follows we put $\hbar=1.$
The eigenenergy is conveniently parameterized by
\be
E=\left [\varepsilon +\alpha-\frac{n_\omega^2}{2}\right]\varepsilon_0.
\label{E-eps}
\ee
where $\varepsilon$ is a dimensionless energy.

\begin{figure}[t]
\centerline{\includegraphics[width=0.3\textwidth]{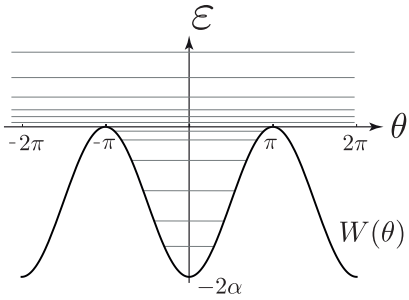} }
\caption{Potential of quantum pendulum.}
\label{pendulum}
\end{figure}

The wave function in the rotating reference frame, $\chi(\theta)$, obeys the differential equation
\begin{equation}
 \label{Shr-stat}
\chi''+2\chi(\varepsilon-W)=0,
\end{equation}
where the double prime stands for the second derivative with respect to the angle $\theta=\phi-\omega t$, while the effective potential is given by
\begin{equation}
\label{pot}
W(\theta)=-\alpha (1+\cos\theta).
\end{equation}
This potential is plotted in Fig.~\ref{pendulum}. The boundary conditions for Eq.~\eqref{Shr-stat} read
\begin{equation}
\chi(0)/\chi(2\pi)=\chi'(0)/\chi'(2\pi)=e^{2\pi i (\phi+n_\omega)}.
\label{bound}
\end{equation}
Thus, in the case of an ideal quantum ring the problem is reduced to the solution of the Schr{\"o}dinger equation which corresponds to a quantum physical pendulum with nonperiodic boundary conditions.

%%%%%%%%%%%%%%%%
%%%%% Fig. 2 pendulum
%%%%%%%%%%%%%%%%
%%%%%%%%%%%%%%%%

\begin{figure}[t]
\centerline{\includegraphics[width=0.45\textwidth]{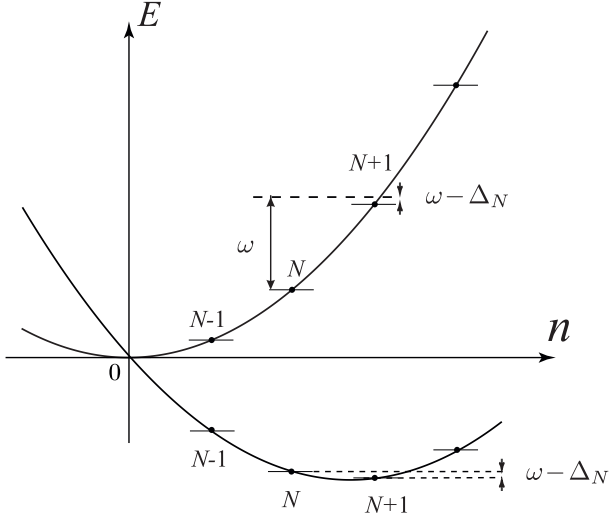} }
\caption{
Energy levels in laboratory and rotation frames.
}
\label{energy}
\end{figure}

The solution to Eq.~\eqref{Shr-stat} supplemented with boundary conditions of Eq.~\eqref{bound} gives rise to the eigenenergies $\varepsilon_n$ and eigenfunctions $\chi_n$.
The corresponding radiation-dressed functions $\Psi_n(\varphi,t)$ are, then, found from Eq.~\eqref{Koshelev2}, where $E=E_n$ is related  to $\varepsilon_n$ by Eq.~\eqref{E-eps}. These functions represent a full basis for electron states in the quantum ring. The energy levels in both laboratory and rotated frame are illustrated in Fig.~\ref{energy}.

%%%%%%%%%%%%%%%%
%%%%% Fig. 3 energy
%%%%%%%%%%%%%%%%
%%%%%%%%%%%%%%%%

The wave-function $\Psi_n$ is also an eigenstate of the current operator which corresponds to the dc current expressed in terms of $\chi_n$ as
\begin{equation}
I_n=I_0 \int {d\theta}\left[\frac{1}{2i}\left(\chi_n^*\frac{\partial \chi_n}{\partial  \theta} -\chi_n \frac{\partial \chi_n^*}{\partial  \theta}\right)+n_{\omega}|\chi_n|^2\right].
\label{Ln}
\end{equation}
In the absence of radiation, i.e.,for $\alpha=0$, we simply obtain
\beml
\label{zerofield}
\begin{eqnarray}
\chi_n &=&\frac{e^{i(n-n_{\omega}-\phi)\theta}}{\sqrt{2\pi}},\quad
\varepsilon_n=\frac{(n-n_{\omega}-\phi)^2}{2},
\label{chi-epsilon- 0}\\
E_n &=& \frac{\varepsilon_0(n-\phi)^2}{2}-\omega(n-\phi).
\end{eqnarray}
\eml
From Eqs.~\eqref{zerofield} it is easy to see that the conventional results for a ballistic single-channel ring are restored in the laboratory frame,
\be
\Psi_n^{(0)}=e^{-iE_n^{(0)}t}\frac{e^{in\varphi}}{\sqrt{2\pi}},\qquad
E_n^{(0)}=\frac{\varepsilon_0(n-\phi)^2}{2}.
\label{E0-Psi0}
\ee
The current corresponding to $n-$th level in the absence of radiation is given by Eq.~\eqref{In}. By weighting these currents with the corresponding Fermi function $f_F \left[E_n^{(0)} \right]$ we arrive with the help of the Poisson summation formula at the expression for persistent current in a clean ring,
\be
I_{\rm{per}} (\phi) =  \sum_{m=-\infty}^\infty I_m \sin(2\pi m\phi),
\ee
where $I_m=-I_0\int dx\;  x \sin(2\pi m x ) f_F(\varepsilon_0x^2/2)$.  One can see that $I_{\rm{per}}(\phi)$ decreases exponentially with temperature [proportional to $\exp(-T/\Delta_F)$] and $I_{\rm{per}} (0)=0$. The total current $I = I_{\rm{per}} + I_{\rm{rad}}$, however, includes an additional radiation-induced contribution $I_{\rm{rad}}$.

\section{Radiation-induced current in the weak coupling  regime}
Let us now turn to the radiation-induced contribution to the current in the clean ring in the case of relatively large temperatures such that
\be
\varepsilon_0 \ll \Delta_F \ll T \ll E_F,
\label{ineq}
\ee
where $\Delta_F \simeq \varepsilon_0 n_F$ is the level spacing at the Fermi level and $n_F \simeq (2E_F/\varepsilon_0)^{1/2} \gg 1$.  In contrast to the persistent current,  the radiation-induced contribution $I_{\rm{rad}}$ is not exponentially suppressed in this regime. Still, similarly to persistent current, $I_{\rm{rad}}$ varies periodically with $\phi$  and, therefore, can be tuned by external field.

We consider first the case of an isolated ring disregarding coupling to the thermal bath. The amplitude of radiation is also assumed to be switched on adiabatically.  Later on, we generalize the obtained result to account for relaxation processes.

\subsection{Isolated ring, adiabatic radiation switching} \label{adiab}
For adiabatic switching a one-to-one correspondence between unperturbed quantum states and radiation-dressed eigenfunctions can be established.  Namely, the   states described by $\Psi_n^{(0)} $ in Eq.~\eqref{E0-Psi0} transform adiabatically into $\Psi_n$. Assuming naturally that the unperturbed system was in a thermal equilibrium we arrive at the following result:
\be
\label{Irad0}
I_{\rm{rad}} = \sum\limits_{n=-\infty}^{\infty}\!\!\delta I_n\, f_n,
\ee
where $\delta I_n=I_n-I_n^{(0)}$, $I_n$ is the current corresponding to radiation dressed functions $\Psi_n$, and $f_n=f_F[E_n^{(0)}]$ is the equilibrium distribution function over unperturbed energy levels. The result of Eq.~\eqref{Irad0} can be rewritten as
\be
I_{\rm{rad}}=\sum\limits_{n=-\infty}^{\infty} J_n \left(f_{n} -f_{n+1} \right),
\label{I-rad}
\ee
where we introduced
\be
J_n=\sum \limits_{m=-\infty}^{n} \delta I_m.
\label{J}
\ee
In the limit of relatively large temperatures, such that inequalities \eqref{ineq} hold, one can further simplify Eq.~\eqref{I-rad} as
\BEA
\label{I-rad1}
 I_{\rm{rad}} &\simeq& -\sum\limits_{n=-\infty}^{\infty} {J_n} \frac{\p f_n}{\p n}
 \\ \nonumber
  &\simeq&   \sum\limits_{n=-\infty}^{\infty} \frac{J_n}{\displaystyle  4  n^* \cosh^2\left[{(n-n_F-\phi)}/{ 2  n^*} \right] },
\EEA
where
\be
n^* = T/\Delta_F \gg 1
\label{n*}
\ee
is the number of quantum levels in the temperature window. The term $\phi/2 n^*$ in the argument of hyperbolic cosine is small but it is needed to preserve the exact invariance of Eq.~\eqref{I-rad1} with respect to a shift of magnetic flux by a flux quantum: $\phi \to \phi+1$.

Let us now evaluate $J_n$ for the case of weak coupling to external radiation, i.e., for  $\alpha \ll 1$,  by taking advantage of perturbation theory. Keeping terms up to the second order with respect to $\alpha$ one finds the spectrum
\be
\varepsilon_n =\frac{(n-n_{\omega}-\phi)^2}{2}-\alpha+\frac{\alpha^2}{4\left[(n-n_{\omega}-\phi)^2-{1}/{4}\right]}
\label{E-pert}
\ee
in accordance with earlier work by Kibis.\cite{Kibis11} We note that optical field induces a change in the current for each radiation-dressed quantum level (this effect was not discussed in Ref.~\onlinecite{Kibis11}). To the second order in $\alpha$ [with the same precision as in Eq.~\eqref{E-pert}] we obtain
\be
\delta I_n=-I_0\frac{\alpha^2}{2}\frac{n-n_{\omega}-\phi}{\left[(n-n_{\omega}-\phi)^2-{1}/{4}\right]^2}.
\label{dIn}
\ee

Substitution of Eq.~\eqref{dIn} into Eq.~\eqref{J} yields
\be
J_n=I_0\frac{\alpha^2}{4}\frac{1}{ (n-n_{\omega}-\phi +1/2)^2} =  I_0\frac{\alpha^2}{4 \delta_n^2},
\label{Jn}
\ee
where we introduced
\beml
\beq
\delta_n &=&  \frac{\omega - \Delta_n}{ \varepsilon_0}=n_{\omega} -n+\phi -1/2,
\label{delta-small-n}\\
 \Delta_n &=&  {E_{n+1}^{(0)}- E_{n}^{(0)}}= {\varepsilon_0(n-\phi+1/2)}.
\label{Delta-n}
\eq
\eml
The energy $ \Delta_n$ is nothing but the spacing between the level $n+1$ and $n$.

It is evident from Eq.~\eqref{Jn} that $J_n$ is strongly enhanced provided a resonance condition $\omega \approx  \Delta_n$ for a given $n$.  Let us assume that such a resonance takes place for $n=N$ such that $\delta_N \ll 1$. In this case one also finds  $|\delta_n| \gtrsim 1$ for all $n \neq N$. Consequently, the current is dominated by the contribution coming from transitions between the levels $N$ and $N+1$.  It is immediately concluded from Eq.~\eqref{Jn} that the perturbation theory applies for $\alpha \ll \delta$, but fails in the opposite limit. Let us, therefore, modify  Eq.~\eqref{Jn} to take into account non-perturbative effects.

\begin{figure}[t]
\centerline{\includegraphics[width=0.4\textwidth]{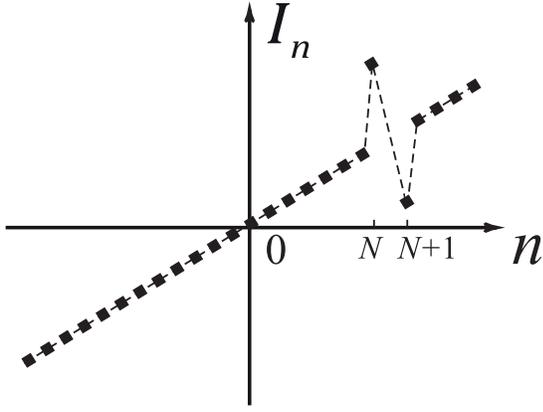} }
\caption{
Currents of quantum levels  for resonant excitation  ($\omega \approx \Delta_N$)   and weak  coupling to radiation  ($\alpha \ll 1$).
}
\label{current-weak}
\end{figure}

In order to evaluate the resonant contribution to the current let us for a moment neglect all optically-induced transitions except  for the transition between levels $N$ and $N+1$. In the rotating wave approximation, the corresponding two-level Hamiltonian reads
\begin{equation}
\hat H = \begin{bmatrix}
\varepsilon_{N}& W\\
W &\varepsilon_{N+1}
\end{bmatrix}
\label{H}
\end{equation}
where the wave functions in the rotation frame  $\chi_N, \chi_{N+1}$ and the corresponding energies  $\varepsilon_N, \varepsilon_{N+1}$ are given by Eqs.~\eqref{chi-epsilon- 0}, while $W$ stands for the matrix element of the optical transition $N+1 \leftrightarrow N$,
\begin{equation}
W= -\alpha \int_0^{2\pi}{e^{i\theta}\cos{(\theta)}d\theta}=-\frac{\alpha}{2}.
\end{equation}
The eigenfunctions of the projected Hamiltonian \eqref{H} are given by
\beml
\label{chiNNN}
\beq
\tilde \chi_N &=& \frac{\chi_N-\beta ~\chi_{N+1}}{\sqrt{1+\beta^2}},
\label{chiN}
\\
\tilde \chi_{N+1} &=& \frac{\chi_{N+1}+\beta~ \chi_{N}}{\sqrt{1+\beta^2}},
\label{chiN+1}
\eq
\eml
where we introduced yet another parameter
\be
\beta=\frac{\alpha\;{\rm sign} (\delta_N) }{|\delta_N|+\sqrt{\delta_N^2+\alpha^2}}.
\label{beta}
\ee
The phase factors in Eqs.~\eqref{chiNNN} are taken in such a way that functions $\tilde{\chi}_N $ and $ \tilde{\chi}_{N+1}$ transform, respectively, into $\chi_N$ and $\chi_{N+1}$  for both positive and negative $\delta_N$ for $\alpha \to 0$.  This leads to appearance of a modulus of $|\delta_N|$ and ${\rm sign}(\delta_N)$ in  Eq.~\eqref{beta}. From Eq.~\eqref{Ln} one obtains the result for currents
\be
\delta I_{N}=-\delta I_{N+1}=I_0 \frac{\beta^2}{1+\beta^2},
\label{INN+1}
\ee
which is illustrated schematically in Fig.~\ref{current-weak}.  One can see from Eq.~\eqref{INN+1} that the current variations $\delta I_{N}$ and $\delta I_{N+1}$ can be as large as $I_0$. On the other hand the contribution of other transitions with $n \neq N$ is suppressed by a small factor $\alpha^2$ and can be estimated as $\alpha^2 I_0$.  Thus, the dominant contribution to Eq.~\eqref{I-rad} indeed comes from the transition with $n=N$. From Eq.~\eqref{INN+1} we find
\be
J_N= \delta I_{N}= \frac{I_0}{2} \frac{\alpha^2}{\sqrt{\alpha^2+\delta_N^2}(|\delta_N|+ \sqrt{\alpha^2+\delta_N^2})}.
\label{JNN}
\ee
Following the procedure described in the beginning of the section we neglect transitions between levels that vary adiabatically and, therefore, refer to the equilibrium distribution of the unperturbed system (i.e.,to the state before the external radiation is adiabatically switched on).  By doing so we arrive at the following expression for the radiation-induced current,  $I_{\rm{rad}}\simeq J_N \left(f_N-f_{N+1}\right)$, for $\omega \approx \Delta_N$:
\be
I_{\rm{rad}} \simeq \frac{J_N}{\displaystyle  4  n^* \cosh^2\left[{(N-n_F-\phi)}/{ 2  n^*} \right]},
\label{I-rad-res}
\ee
where we take into account  that in the adiabatical case the current $\delta I_N$ must be weighted with the unperturbed Fermi distribution function $f_n$.

\begin{figure}[t]
\centerline{\includegraphics[width=0.3\textwidth]{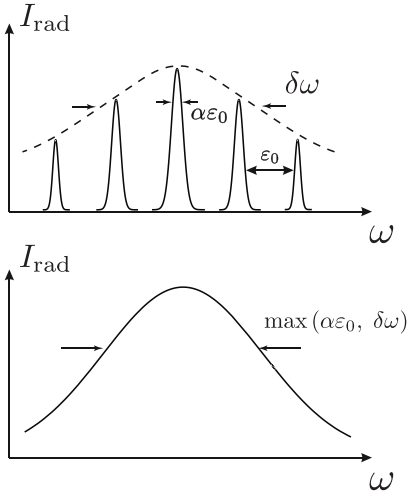} }
\caption{{Upper panel}: Radiation-induced circulating current for the case of   weak coupling to radiation.   The resonances corresponding to excitations of different pairs of levels are well separated;     {Lower panel}: Resonances overlap  and    merge into a single wide peak for the case of strong coupling.
   }
\label{current-both}
\end{figure}

Summing up the contributions from all levels we arrive at a more general result which includes both non-resonant and resonant contributions
\beq \nonumber
I_{\rm{rad}} &\simeq& \frac{I_0}{2} \sum\limits_{n=-\infty}^{n=\infty}
\frac{\alpha^2}{\sqrt{\alpha^2+\delta_n^2}(|\delta_n|+ \sqrt{\alpha^2+\delta_n^2})} \\ &\times&   \frac{1}{\displaystyle  4  n^* \cosh^2\left[{(n-n_F-\phi)}/{ 2  n^*} \right] }.
\label{I-rad-final}
\eq
The dependence of the current $I_{\rm{rad}}$ given by Eq.~\eqref{I-rad-final} on frequency is shown in the upper panel of Fig.~\ref{current-both}. Positions of the peaks are found from the conditions $\delta_n=0.$  The smooth envelope of the peaks is due to the  thermal factor $ {1}/{\displaystyle  4  n^* \cosh^2\left[{(n-n_F-\phi)}/{ 2  n^*} \right]}.$ The central peak corresponds to resonance excitation of levels at the Fermi energy while its amplitude corresponds to a maximal possible optical response for weak coupling, which can be estimated as
\be
I_{\rm{rad}}^{\rm max} \simeq \frac{I_0}{8n^*}=\frac{I_0 \Delta_F}{8T}.
\label{Imax-weak}
\ee
The distance between resonant peaks is given by $\varepsilon_0$. Remarkably, the coupling strength $\alpha$ drops out from the result of Eq.~\eqref{Imax-weak}.  Thus, the radiation-induced contribution to current might be large even in the weak-coupling regime, and decays as $T^{-1}$ in contrast to the persistent current contribution which decays exponentially with temperature.

It is worth stressing  that positions of resonances depend on $\phi$ and are therefore tunable by magnetic flux piercing the ring.  Changing the magnetic flux by the flux quantum, $\phi\to\phi+1$ is equivalent to the substitution $\delta_n\to\delta_{n-1}$, which proves that the result of Eq.~\eqref{I-rad-final} is a periodic function of $\phi$ with the period $1$.  Thus, instead of varying frequency of radiation one can probe optically-driven resonances in the ring by varying external magnetic field.

Let us  present analytical expression for the smooth envelope $I_{\rm env} (\omega)$ of the resonance peaks plotted with the dashed line in the upper panel in Fig.~\ref{current-both}. The maximal peak values are found from the condition  $\delta_n=0$, which is equivalent to $n \approx n_\omega +\phi-1/2$.  Substituting the latter equality in the thermal factor in Eq.~\eqref{I-rad-final} one finds
\beq
I_{\rm env} (\omega) &\approx &\frac{I_0}{8n^*}~\frac{1}{\cosh^2[(n_\omega-n_F -1/2)/2n^*]}  \nonumber \\
&\approx &\frac{I_{\rm{rad}}^{\rm max}}{\cosh^2[(\omega-\Delta_F )/2\delta\omega]},
\label{env}
\eq
where $\delta \omega  =\varepsilon_0 n^*=T/ n_F$ is the envelope width. From the physics point of view the frequency range $\delta \omega$ corresponds to the variation of the level spacing within the temperature window, hence $\delta \omega \simeq (\p \Delta_n/ \p n) n^*$.

It is worth noting that Fig.~\ref{current-both} corresponds to the case of positive helicity ($\omega >0$). The current changes sign under simultaneous inversion of helicity and magnetic field [see Eq.~\eqref{Irad_phi_omega}], i.e.,under the time reversion. Consequently, the field-independent contribution to the current changes sign under the inversion of helicity only.

Before closing this subsection, we shall briefly discuss the role of mesoscopic fluctuations.  As we mentioned above, such fluctuations dominate persistent-current especially at high temperatures, when the averaged value of $I_{\rm per}$ is exponentially small. Let us estimate the fluctuations of $I_{\rm rad}$ using simple arguments. First of all we shall notice that only the electron levels within the temperature window around the Fermi level can be populated or depopulated.  The number of such levels is  estimated as $n^*$ [see Eq.~\eqref{n*}]. Thus, the fluctuations of the total number of electrons in the ring are given by $\Delta N \sim\sqrt{n^*} \sim \sqrt{T/\Delta_F}$. The corresponding fluctuation of the chemical potential reads $\Delta \mu = \Delta_F \delta N \sim  \sqrt{T \Delta_F}$. Such fluctuations do not affect our results for $I_{\rm per}$ provided the factor ${\p f_n}/{\p n}$ entering Eq.~\eqref{I-rad1} does not fluctuate much. This is indeed the case for $\Delta \mu/T \ll 1$.  In this limit we estimate the mesoscopic fluctuation of the optically-induced current as
\be
\frac{\Delta I_{\rm rad}}{ I_{\rm rad}} \sim \frac{\Delta \mu}{T} \sim    \sqrt{\frac{\Delta_F}{T} } \ll 1.
\ee

Thus, we conclude that for high temperatures mesoscopic fluctuations of $I_{\rm rad}$ are suppressed.\cite{linear} This, in turn, implies that, in contrast to persistent current,  $I_{\rm rad}$ is not very sensitive to the choice of thermodynamic statistical ensemble. The robustness of $I_{\rm rad}$ with respect to the mesoscopic fluctuations can be understood rather easily from a simple physics argument. Indeed, the resonant condition for a pair of neighboring  levels in the temperature window near the Fermi surface is satisfied provided that interlevel distance equals the radiation frequency. Since the distance between levels does not depend on ensemble, the only  condition required is that the fluctuations do not move the majority of relevant levels out of the temperature window. This condition is equivalent to $T\gg \Delta \mu$.
%%%%%%%%%%%%%%%%
%%%%% Fig. 4 current-weak
%%%%%%%%%%%%%%%%
%%%%%%%%%%%%%%%%

%%%%%%%%%%%%%%%%
%%%%% Fig. 5 current-both
%%%%%%%%%%%%%%%%
 %%%%%%%%%%%%%%%%

%%%%%%%%%%%%%%%%
%%%%% Fig. 6 current-strong
%%%%%%%%%%%%%%%%
\begin{figure}[t]
\centerline{\includegraphics[width=0.4\textwidth]{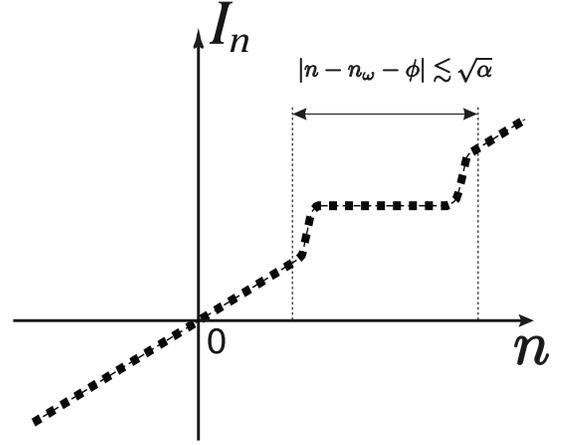} }
\caption{Currents of quantum levels  for strong  coupling to radiation  ($\alpha \gg 1$).
   }
\label{current-strong}
\end{figure}
%%%%%%%%%%%%%%%%

\subsection{Ring coupled to thermal bath} \label{bath}
In this subsection we shall turn to the case of ideal quantum ring coupled in addition to a thermal bath while still assuming $\alpha \ll 1$, i.e.\,a weak coupling to electromagnetic radiation. We will see that in this case qualitative physical picture drown in Fig.~\ref{current-both} and in previous subsection remains intact provided that the rate of relaxation is sufficiently low. The only essential difference is related to the shape of the optically-induced resonances in the current.

Similarly to the previous section we assume $\delta_N\ll 1$ and consider the only pair of resonant levels $N$ and $N+1$.  In this regime the effect of the thermal bath can be described by the simplest Markovian model, which is expressed in terms of the density matrix for the two-level system as
\beml
\label{mtrx}
\beq
\dot\rho_N &=&- \alpha\varepsilon_0\;{\rm Im}[e^{i\omega t}\rho] + \gamma ( f_N-\rho_N),\\
\dot\rho_{N+1} &=&\alpha\varepsilon_0\; {\rm Im}[e^{i\omega t}\rho]+\gamma( f_{N+1}-\rho_{N+1}),\\
-\dot\rho &=&(i\Delta_N+\gamma_\varphi)\rho+\frac{i\alpha \varepsilon_0}{2} (\rho_{N+1}-\rho_N)e^{-i\omega t},\qquad
\eq
\eml
where $\rho_{N+1}=\rho_{N+1,N+1}$, $\rho_{N}=\rho_{N,N}$, and $\rho=\rho_{N+1,N}$ parametrize the relevant components of the density matrix, while $\gamma$ and $\gamma_\varphi$ are the relaxation and dephasing rates, respectively. In what follows we substitute
\begin{equation}
\rho=e^{-i\omega t}\tilde \rho
\end{equation}
and search for stationary solutions of Eqs.~\eqref{mtrx}. Straightforward analysis yields
\beml
\label{drhoN}
\beq
\delta \rho_N &=&
\frac{\alpha^2\Gamma_\varphi (f_{N+1}- f_N)}{2[\Gamma (\delta_N^2+\Gamma_\varphi^2) +\alpha^2\Gamma_\varphi]}= - \delta\rho_{N+1},\\
\tilde{\rho} &=& \frac{\alpha (\rho_{N+1}-\rho_{N})}{2(\delta_N+i\Gamma_\varphi)},
\eq
\eml
where $\delta\rho_{m}=\rho_{m} - f_N$ is the radiation-induced variation of level population while $\Gamma= \gamma/\varepsilon_0$ and $\Gamma_\varphi= \gamma_\varphi/\varepsilon_0$ stand for dimensionless rates. The corresponding expression for dc current reads
\begin{equation}
I_{\rm rad}=I_0\sum_{n=-\infty}^{\infty}{(n-\phi)~\delta\rho_{n,n}}.
\label{I-rad-bath0}
\end{equation}
Substituting  Eq~\eqref{drhoN} into  Eq.~\eqref{I-rad-bath0}, we obtain the result
\begin{equation}
I_{\rm rad}=\frac{I_0}{2}\frac{\alpha^2\Gamma_\varphi (f_{N}- f_{N+1})}{\Gamma (\delta_N^2+\Gamma_\varphi^2) +\alpha^2\Gamma_\varphi}.
\label{I-rad-bath}
\end{equation}
Thus, the resonant radiation-induced current in the presence of a thermal bath is still given by Eq.~\eqref{I-rad-res} with
\be
J_N=\frac{I_0}{2}\frac{\alpha^2\Gamma_\varphi }{\Gamma (\delta_N^2+\Gamma_\varphi^2) +\alpha^2\Gamma_\varphi}.
\label{J_Ndeph}
\ee
The inequality $\Gamma_\varphi  \geq \Gamma/2$  holds since relaxation of level populations leads to dephasing. Assuming that there are no other sources of dephasing one can simplify Eq.~\eqref{J_Ndeph} to
\be
J_N=\frac{I_0}{4}\frac{\alpha^2 }{ \delta_N^2+\Gamma^2/4 +\alpha^2/2}.
\label{J_Nnodeph}
\ee
By comparing Eq.~\eqref{J_Nnodeph} with Eq.~\eqref{JNN} we conclude that the maximal value of $J_N$  is smaller for the ring coupled to a thermal bath than for an isolated ring by a factor $\alpha^2/(\alpha^2+\Gamma^2/2)$ as intuitively expected. That explains why the resonance width increases from $\alpha$ to $ \sqrt{\alpha^2+ \Gamma^2/2}$.

For small relaxation rate, $\Gamma \ll 1$, the  dc current shows a series of sharp resonances in full analogy to the case of isolated adiabatic system. In the limit $\Gamma \to 0$  the only difference between these two cases is related to the different shapes of resonances.  For a finite $\Gamma$  the difference becomes more essential due to the presence of some dissipation caused by interaction with the bath. The dissipated power $P$ can be estimated from the conventional formula for Joule heating
\be
P = R\int d\varphi\left \langle I(\varphi,t) \mathcal E_\varphi (\varphi, t)  \right \rangle_t,
\label{P}
\ee
where $\mathcal E_\varphi (\varphi, t)=-\mathcal E_0\sin(\varphi-\omega t)$ is the projection of the electric field on the current direction, $R$ is the ring radius, and the current is given by a generalization of Eq.~\eqref{current00}  for the system described with the help of the density matrix,
\be
I(\varphi,t)=I_0\sum\limits_{n,m}\rho_{nm}(t) \left(\frac{n+m}{2} -\phi\right)e^{i(n-m)\varphi}.
\label{Inm}
\ee
In contrast to Eq.~\eqref{current00} we, however, should keep in Eq.~\eqref{Inm}  the ac contribution to the current. It is precisely this contribution that is responsible for dissipation. Straightforward analysis yields the dissipated power
\be
P=2\pi R\mathcal E_0 I_0   \frac{\alpha\Gamma}{4}  \frac{(N+1/2-\phi)(f_N-f_{N+1})}{\delta_N^2+\Gamma^2/4+\alpha^2/2},
\ee
which is proportional to both the coupling constant $\alpha$ and the rate $\Gamma$ which characterizes the coupling to the bath.

\section{Strong coupling  to the radiation} \label{strong}

\subsection{Isolated ring, adiabatic radiation switching}
In this section we focus on the strong-coupling regime $\alpha \gg 1$. Similarly to the previous section we consider first the case of a completely isolated ring assuming that radiation switches on adiabatically so that we can deduce the level occupation numbers from an equilibrium state at an initial moment of time. We, then, turn to the case of a ring coupled to a thermal bath for which we do not need to make such an assumption.

We recall that resonances obtained in the weak-coupling regime (see upper panel of Fig.~\ref{current-both}) are separated by the distance $\varepsilon_0$  and have the width $\alpha \varepsilon_0$.  Therefore, the resonances are expected to overlap if coupling to radiation increases. In the regime of overlapping resonances the two-level approximation used in the previous section is no longer justified and a more accurate analysis has to be performed.  Simple consideration below shows, indeed, that for $\alpha  \gg 1$  the dependence of radiation-induced current on frequency is given by a single peak of a large amplitude. This dependence is depicted schematically in the lower panel of Fig.~ \ref{current-both}.

For an isolated ring strongly coupled to radiation the effective potential $W(\theta)$ is large enough to localize the states near $\theta =0$ in the rotating frame.  These localized states correspond to the energy range $\varepsilon<0$ in  Fig.~\ref{pendulum}.  Simple quasiclassical analysis of Eq.~\eqref{Shr-stat}  shows that the total number of localized states is of the order of $\sqrt{\alpha}$ while the distance between the levels is proportional to $\sqrt{\alpha} \varepsilon_0$.  These energies are close to the bottom of the parabola in Fig.~\ref{pendulum} and correspond to $n$ lying in a vicinity of $n_\omega$ such that $|n-n_\omega-\phi | \lesssim \sqrt{\alpha}$. In the laboratory frame the localized states form a band of the width $\sqrt{\alpha}$ centered around $n = n_\omega$.  In the absence of disorder all states in the band  have a certain chirality.  The helicity of the radiation determines the sign of $n_\omega$ and, consequently, the chirality of the localized band.

Since wave functions $\chi_n(\theta)$ for localized states are real the only term which contributes to the radiation-induced current is the last one in Eq.~\eqref{Ln}. Using the normalization  condition $\int |\chi_n(\theta)|^2d\theta=1$ we estimate the localized state contributions to the current $I_n \approx I_0 n_\omega  = \textrm{constant}$  for $|n-n_\omega-\phi| \lesssim \sqrt{\alpha}|$ as illustrated in Fig.~\ref{current-strong}. On the other hand for energies outside the localized  band, i.e.,for  $|n-n_\omega-\phi| \gg \sqrt{\alpha}$, the radiation does not affect the current in any essential way hence the perturbative result of Eq.~\eqref{dIn} remains valid. With the help of Eq.~\eqref{J} we obtain
\be
J_n  \approx I_0
\begin{cases}
\frac{\alpha^2}{4(n-n_\omega -\phi)^2},& | n\!-\!n_\omega\! -\!\phi| \gg\sqrt \alpha,\\
C_1 \alpha-\frac{(n-n_\omega -\phi)^2}{2},& | n\!-\!n_\omega\! -\!\phi| \ll \sqrt \alpha,
\end{cases}
\label{Jn1}
\ee
where $C_1\sim 1$ is a numerical coefficient.  Let us now assume that the temperature is sufficiently large so that $n^* \gg \sqrt \alpha$, or, equivalently, $T\gg \sqrt \alpha \Delta_F$. Then, the distribution function does not change within the width of the band. Substituting Eq.~\eqref{Jn1} into  Eq.~\eqref{I-rad1} and replacing  the summation over $n$ with integration we estimate the current as
\be
I_{\rm rad} \sim  \frac{I_0\alpha^{3/2}}{n^*\cosh^2[(n_\omega-n_F)/2n^*] }.
\label{I-rad-strong}
\ee
This result suggests that the maximal value of the current
\be
I_{\rm rad}^{\rm max} \sim  \frac{I_0\alpha^{3/2}}{n^*}.
\ee
is achieved for $\omega \approx \Delta_F$ while the width of the broadened resonant peak is proportional to $T/ n_F$.  Comparing this result with that of Eq.~\eqref{Imax-weak} we conclude that the maximal achievable current increases with the radiation strength.

For sufficiently large values of $\alpha$, such that $T\ll \sqrt \alpha \Delta_F$, the dependence of the derivative $\p f_n/\p n$ on $n$ becomes stronger than that of $J_n$. In this case the current is given by $I_{\rm{rad}} \approx J_{n_F}$, where $J_n$ is still determined from Eq.~\eqref{Jn1}. The maximal value of the current in this case reads
\be
\label{reg1}
I_{\rm rad}^{\rm max} \sim  I_0\alpha,  \quad \text{for}\quad n_F \gg  \sqrt \alpha \gg n^*.
\ee
In the limit of very large coupling  $\sqrt \alpha \gg n_F$, the maximal current saturates at the value
\be
\label{reg2}
I_{\rm rad}^{\rm max} \sim  I_0 n_F^2,    \quad \text{for}\quad      \sqrt \alpha \gg n_F,
\ee
while the width of the peak in both regimes \eqref{reg1} and \eqref{reg2} is given by $\sqrt \alpha \varepsilon_0$.  Equation \eqref{reg2} has a very clear physical sense. For such a large coupling all electrons  are localized in the rotation frame  and rotate with the velocity $v_F \propto n_F.$  The current is given by the ratio of the total charge $Q \simeq e n_F$ to  the time of the electron traveling around the ring,  which is   proportional to $1/v_F.$  This yields Eq.~\eqref{reg2}.

\subsection{Ring coupled to thermal bath}

Similarly to the previous section the radiation-induced current in the presence of the coupling to a thermal bath can be calculated using Eq.~\eqref{I-rad-bath0}. The two-level approximation used above is, however, no longer justified due to a strong overlap of resonances corresponding to optical transitions between particular energy levels.
In this regime the relaxation of diagonal and off-diagonal elements of the density matrix is governed respectively by the terms $\gamma (f_n- \rho_{n,n})$  and  $-\gamma_\varphi  \rho_{n,m}$ in the collision integral. For simplicity we shall neglect the possible dependence of collision rates $\gamma$ and $\gamma_\varphi$ on energy.

The equation on the density matrix takes the form
\beq
\label{F}
 \frac{\p F_n}{\p t} &=&\left[{(n-\phi)\varepsilon_0}- \omega  \right]\frac{\p F_n}{\p\theta} -\frac{i\varepsilon_0}{2} \frac{\p^2 F_n}{\p \theta^2 }  \nonumber\\
&+&\frac{i\alpha \varepsilon_0}{2 } \left( F_{n+1}  e^{i\theta} + F_{n-1} e^{-i\theta}- 2F_{n}\cos\theta \right) \nonumber \\
&+&  \gamma ( f_n- \overline{F}_n) + \gamma_\varphi (\overline{F}_n- {F_n}),
\eq
where $\overline{F}_n= \rho_{n,n}=\tfrac{1}{2\pi}\int_0^{2\pi}\!\! d\theta\, F_n(\theta)$ and
\be
F_n(t, \theta) =  \sum \limits_k \rho_{n,n+k} e^{ik(\theta- \omega t)}.
\ee
The Equation~\eqref{F} is easily analyzed provided $\gamma \gg \varepsilon_0$, $\gamma_\varphi \gg \varepsilon_0$. In this limit we search for a stationary solution to Eq.~\eqref{F} in the following form
\be
F_n=\overline{F}_n + \alpha_n e^{i\theta}+\beta_n e^{-i\theta},
\label{FF}
\ee
where higher harmonics with respect to the angle $\theta$ are neglected.  Substituting Eq.~\eqref{FF} into Eq.~\eqref{F}  we find
\be
\alpha_n = \beta_{n+1}^*=\frac{\alpha \varepsilon_0}{2} \frac{\overline{F}_{n+1}-\overline{F}_n}{\delta_n-i\gamma_\varphi},
\ee
where the function $\overline{F}_n$ is determined by the balance equation
\be
\frac{\alpha^2  \Gamma_\varphi}{2} \left (  \frac{\overline{F}_{n+1}-\overline{F}_n}{\delta_n^2 + \Gamma_\varphi^2}  +\frac{\overline{F}_{n-1}-\overline{F}_{n}}{\delta_{n-1}^2 + \Gamma_\varphi^2}\right)=\Gamma  (\overline{F}_n-f_n),
\label{balance}
\ee
with the same definition of dimensionless collision rates $\Gamma=\gamma/\varepsilon_0$ and $\Gamma_\varphi=\gamma_\varphi/\varepsilon_0$ as in the previous section. Since both collision rates are large $\Gamma \gg 1$, $\Gamma_\varphi \gg 1$ Eq.~\eqref{balance} can be rewritten in the differential form as
\be
\frac{\alpha^2  \Gamma_\varphi }{2\Gamma} \frac{\p}{\p n}\left (  \frac{1}{\delta_n^2 + \Gamma_\varphi^2}  \frac{\p\overline{F}_{n}}{\p n}\right)=\overline{F}_n- f_n.
\label{balance1}
\ee
The solution to Eq.~\eqref{balance1} can be found using the ansatz
\be
\overline{F}_n =f_n+ \Gamma_\varphi G_n \frac{\p f_n}{\p n},
\label{G}
\ee
which is justified for sufficiently high temperatures such that $n^* \gg  \Gamma_\varphi$. Substituting Eq.~\eqref{G} into Eq.~\eqref{balance1} and neglecting terms which are proportional to ${\p^2 f_n}/{\p n^2}$ and ${\p^3 f_n}/{\p n^3}$ we arrive at the following equation for $G=G_n$:
\be
\eta G =\frac{\p}{ \p x} \left[ \frac{1}{1+x^2} \left( 1+\frac{\p G}{ \p x}\right)\right],
\label{GG}
\ee
where $x={\delta_n}/{\Gamma_\varphi}$ and $\eta=2\Gamma \Gamma_\varphi^3/\alpha^2$. The dimensionless parameter $\eta$ characterizes the strength of thermalization rates relative to optical transition rate and can, therefore, be regarded as a measure of thermalization intensity. For $\eta \gg 1$ one can neglect the term $\p G/\p x$ in the right hand side of Eq.~\eqref{GG}.  Thus, for relatively fast thermalization we obtain
\be
G\approx - \frac{2x}{\eta(1+x^2)^2},\quad \text{for}\;\eta \gg 1.
\label{GGG}
\ee
This result also applies for $|x|\gg \eta^{-1/4}$ irrespective of the value of $\eta$. For $|x|\ll \eta ^{-1/4}$ one can disregard the left hand side of Eq.~\eqref{GG}. Thus, the behavior of $G$ in the limit of relatively slow thermalization is given by
\be
G \approx
\begin{cases}
 -x,&   |x| \ll \eta ^{-1/4},\\
- \frac{2}{\eta x^3},& |x| \gg\eta ^{-1/4},
\end{cases}\quad \text{for}\;\eta \ll 1.
\label{GGGG}
\ee
The results of Eqs.~\eqref{GGG} and \eqref{GGGG} have to be substituted into Eq.~\eqref{G} in order to obtain the distribution function. Using the latter in Eq.~\eqref{I-rad-bath0}, taking into account that $\overline{F}_n= \rho_{n,n},$  and replacing the summation over $n$ with integration we arrive at the following result for current
\be
I_{\rm rad} \simeq \frac{I_0}{n^*\cosh^2\left[\frac{n_\omega-n_F}{2n^*}\right] }
\begin{cases}
\alpha^{2}/\Gamma, & \alpha^2 \ll   \Gamma_\varphi^{3} \Gamma, \\
\alpha^{3/2}(\Gamma_\varphi/\Gamma)^{3/4}, & \alpha^2 \gg  \Gamma_\varphi^{3} \Gamma.
\end{cases}
 \label{I-rad-strong1}
 \ee
It is worth noting that in the limit $\Gamma_\varphi \sim \Gamma $ and $\alpha \gg \Gamma^2$  the second line in Eq.~\eqref{I-rad-strong1} coincides with the result of Eq.~\eqref{I-rad-strong} obtained for the adiabatic case.  We should also note that in the derivation of the asymptotic behavior for large $\alpha$ expressed by the second line in Eq.~\eqref{I-rad-strong1} the large-temperature limit, $T\gg \Delta_F\sqrt{\alpha}$, has been implicitly assumed.

\section{Disordered ring} \label{disorder}

Surface roughness, impurities and external Coulomb potentials are the main sources of disorder which are almost impossible to avoid in a realistic nanoring. The disorder leads to backscattering that would invalidate the analysis of the previous sections. In this paper we do not consider the limit of strong disorder such that Anderson localisation on the scale of the nanoring circumference sets in.  Instead we focus on the cleanest possible but still realistic systems for which disorder can be regarded as small, i.e.,the corresponding mean-free path is large or comparable with the ring radius. We further distinguish the cases of short-range and long-range disorder.

The short-range disorder leads to a scattering between right- and left- moving electrons. One may naively expect that such processes would merely lead to additional broadening of the resonant peaks for the radiation-induced current. Contrary to expectations the weak short-range disorder is shown below to split the resonances and induce new narrow resonance peaks with the amplitude enhanced by a large factor of the order of $n_F$ as compared to that in the case of an ideal ring. The suppression of these new resonant features happens only for sufficiently strong disorder.

The effect of long-range smooth disorder is entirely different.  A long-range potential does not affect the result for radiation-induced current in the limit of weak coupling $\alpha\ll 1$, but may affect optical transitions if the light amplitude is sufficiently large. In the latter case the physics of the system is equivalent to those of a physical pendulum described within the quasiclassical approximation. Random long-range potential leads to a classical chaotic behavior of the system, which results in the appearance of a thin chaotic layer near the separatrix of the physical pendulum.

\subsection{Weak short-range disorder}

Let us start with a more detailed analysis of the model in the presence of weak short-range disorder. In an ideal ring all energy levels are double degenerate provided the dimensionless magnetic flux $\phi$ is an integer or half-integer number.  In both cases every level with a positive chirality has a partner with a negative chirality which corresponds to the same energy. A short-range disorder induces backscattering which prevents the use of chirality as a quantum number and mixes the pairs of degenerate states.

For the sake of definiteness we shall focus on the vicinity of $\phi= 0$ first. We further assume that $\omega \approx \Delta_N$ and  $\alpha \ll 1$. One may still remember from the analysis of the previous section that under such conditions the resonant optical transition between $N-$th and $(N+1)-$th level is the only one which is relevant in an ideal ring. The presence of disorder potential, $U(\varphi)$, mixes the states of positive and negative chiralities. If both disorder and radiation are sufficiently weak (the corresponding conditions will be formulated below)  the resonances in radiation-induced current can be obtained within a 4-level approximation based on the Hamiltonian projected on the states $\pm N$ and $\pm(N+1)$,
\be
\hat H =
\begin{pmatrix}
E_{N}^{(0)} & \alpha \varepsilon_0 e^{i\omega t}/2 & U_N^* & 0\\
\alpha \varepsilon_0 e^{-i\omega t}/2 & E_{N+1}^{(0)} & 0 & U_{N+1}^ *\\
U_N & 0 & E_{-N}^{(0)} & 0 \\
0 & U_{N+1} & 0 & E_{-(N+1)}^{(0)} \\
\end{pmatrix},
\label{H4}
\ee
where
\be
\label{UNN}
U_N= \frac{1}{2\pi} \int  U(\varphi)  \, e^{-2 i N\varphi}{d\varphi}
\ee
is the matrix element of the disorder potential $U(\varphi)$ mixing the states $N$ and $-N$. In the effective model of Eq.~\eqref{H4} we still neglect the matrix element corresponding to optical transition between the levels $-N$ and $-N-1$ by keeping in mind that such a transition is non-resonant for circularly polarized light.

It is instructive to diagonalize the effective Hamiltonian \eqref{H4} with respect to disorder potential. The corresponding basis states are, then, conveniently numerated by the index $L$ or $R$ and by $n=N$ or $N+1$,
\beml
\beq
\Psi_n^L=\frac{\xi_n^*e^{i n \varphi}+ e^{-in\varphi}}{\sqrt{2\pi(1+|\xi_n|^2)}},\quad
\Psi_n^R=\frac{e^{i n \varphi}- \xi_ne^{-in\varphi}}{\sqrt{2\pi(1+|\xi_n|^2)}},&&\quad\label{psi}\\
E_n^{L(R)} =D_n \pm \sqrt{D_n^2 +|U_n|^2},&& \label{E-RL}
\eq
\eml
where the following notations are introduced
\beml
\beq
D_n &=&\frac{E_{-n}^{(0)} -E_{n}^{(0)}}{2}=\phi\, n \varepsilon_0, \label{Dn}\\
\xi_n &=&\frac{U_n}{D_n +\sqrt{|U_n|^2+D_n^2}}.
\eq
\eml
For $U_n=0$, the functions $\Psi_n^R$ and  $\Psi_n^L$ correspond to right- and left-moving states, $\exp(i n \varphi)$ and $\exp(-in\varphi),$ respectively.

Using the basis functions $\Psi_N^R$, $e^{-i\omega t}\Psi_{N+1}^R$, $\Psi_N^L$,  and $e^{-i\omega t} \Psi_{N+1}^L$,  we rewrite the Hamiltonian given by Eq.~\eqref{H4} in manifestly time-independent form,
\be
\hat H'\! =\!
\begin{pmatrix}
E_{N}^{-} &  V &  0 &  V\xi^*_{N+1}\\
V &  E_{N\!+\!1}^{-} \!-\!\omega &  V\xi^*_{N}  &   0\\
0 &  V\xi_{N} &  E_{N}^{+} & V\xi_{N}\xi^*_{N+1} \\
V\xi_{N+1} &  0 &  V\xi_{N}^*\xi_{N+1}  &  E_{N\!+\!1}^{+}\!-\!\omega \\
\end{pmatrix},
\label{H4-prime}
\ee
where
\be
V= \frac{\alpha \varepsilon_0}{2\sqrt{(1+|\xi_N|^2)(1+|\xi_{N+1}|^2)}}.
\label{V}
\ee

In the absence of disorder one has $\xi_N=\xi_{N+1}=0$, hence the only optical transition allowed is the $RR$ transition between the states $N+1$ and $N$, which both correspond to right-moving electrons. The corresponding matrix element equals $\alpha \varepsilon_0/2 = e\mathcal{E}_0R/2$.

\begin{figure}[t]
\centerline{\includegraphics[width=0.350\textwidth]{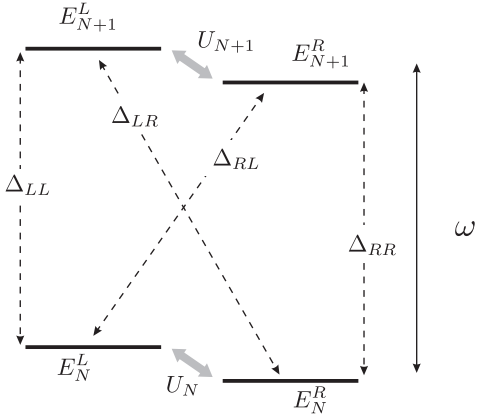} }
\caption{Resonant ($\omega \approx \Delta_N$) optical transitions between levels of right- and left- moving electrons. Transitions $LL,$ $RL,$ and $LR$    are induced by disorder.}
\label{levels}
\end{figure}

In the presence of weak short-range disorder the original states for right- and  left-moving electrons are mixed. As a result the $RR$ transition occurs between the states $\Psi_{N+1}^R$ and $\Psi_{N}^R$ with the corresponding matrix element $V_{RR}=V$. In addition three more transitions emerge $\Psi_{N+1}^L \!\!\!\leftrightarrow \!\! \Psi_N^R$,  $\Psi_{N+1}^R  \!\!\! \leftrightarrow\!\!  \Psi_N^L$, and $\Psi_{N+1}^L \!\!\!\leftrightarrow \!\!  \Psi_N^L$ that are labeled as $LR$,  $RL$,  and  $LL$, respectively (see   Fig.~\ref{levels}). As can be seen from Eq.~(\ref{H4-prime}) the matrix elements corresponding to these transitions are $V_{LR}=V\xi_{N+1}$, $V_{RL}=V\xi_{N}^*$,  and  $V_{LL}=V\xi_{N}^*\xi_{N+1}$. The corresponding resonant frequencies are given by
\be
\Delta_{ab}=E^a_{N+1}- E^b_N,
\ee
where $E_n^{a}$ is given by Eq.~\eqref{E-RL} and the indices $a,b$ take on $R,L$.
%%%%%%%%%%%%%%%%
%%%%% Fig. 7  levels
%%%%%%%%%%%%%%%%
%%%%%%%%%%%%%%%%
For small $V$ all four optical transitions are well resolved. If radiation frequency is close to one of the resonant frequencies $\Delta_{ab}$ one may again use a two-level  approximation described by the effective Hamiltonian
\be
\hat H_{ab} =
\begin{pmatrix}
E_{N}^{a}  & V_{ab}^*\\
V_{ab}& E_{N+1}^{b}-\omega \\
\end{pmatrix},
\label{H-LR}
\ee
which acts in the space spanned by the functions $\Psi_{N}^{a}$ and $e^{-i\omega t}\Psi_{N+1}^{b}$. The eigenfunctions of the Hamiltonian \eqref{H-LR} are readily found as
\beml
\beq
\tilde \Psi_N^a &=&\frac{\Psi_N^a-\beta_{ab} e^{-i\omega t}\Psi_{N+1}^b}{\sqrt{1+|\beta_{ab}|^2}}, \label{psi-tilde-N}\\
\tilde \Psi_{N+1}^b &=& \frac{\beta_{ab}^*\Psi_{N}^a + e^{-i\omega t} \Psi_{N+1}^b}{\sqrt{1+|\beta_{ab}|^2}},
\label{psi-tilde-N+1}
\eq
\eml
where we introduce
\be
\beta_{ab}=\frac{2V_{ab}\;\textrm{sign} (\delta_{ab})}{|\delta_{ab}| +\sqrt{\delta_{ab}^2 +4|V_{ab}|^2}},
\quad \delta_{ab}=\omega-\Delta_{ab}.
\ee
The result of Eq.~\eqref{psi} can now be substituted into Eqs.~\eqref{current00}, \eqref{psi-tilde-N},  and \eqref{psi-tilde-N+1} in order to calculate the radiation-induced current. The calculation in the adiabatic case is very similar to those presented in Sec.~\ref{adiab}.  For disorder-induced splitting that is small compared to temperature one finds that the resonant radiation-induced current for $\omega \approx \Delta_N$ is still given by Eq.~\eqref{I-rad-res} with
\be
J_N=I_0\sum_{a,b} \frac{|\beta_{ab}|^2}{1+|\beta_{ab}|^2} \left[ (N+1/2) A_{ab} +B_{ab}/2 \right],
\label{JNNN}
\ee
where we introduced
\beml
\beq
A_{RR}=- A_{LL}= B_{RL}=- B_{LR}= \lambda_{N+1}-\lambda_{N},&&\qquad  \\
B_{RR}=-  B_{LL}= A_{RL}= - A_{LR}= \lambda_N+\lambda_{N+1},\eq
\eml
and
\be
\lambda_n=\frac{1-|\xi_n|^2}{ 1+ |\xi_n|^2}.
\ee
The derivation of Eq.~\eqref{JNNN}, which describes $4$ resonances at frequencies $\Delta_{ab}$, has been based on several important assumptions.

First of all the $4$-level approximation used to justify Eq.~(\ref{H4}) is valid if disorder mixes nearly degenerate levels $n$ and $-n$ only, i.e.,those which have opposite chiralities in the absence of disorder potential. These levels are separated by energy $D_n$ defined in Eq.~\eqref{Dn}, hence the mixing is controlled by the parameters  $U_n/D_n  =  U_n/ n\phi \varepsilon_0$ for $n=N, N+1$.  The admixture of other levels is weak as far as $U_n \ll n \varepsilon_0$, which is the central condition for the validity of Eq.~(\ref{H4}). Note, however, that the relation between $U_n$ and $D_n$ can be arbitrary.

It has been also implied that the resonance frequencies $\Delta_{ab}$ arising due to the splitting of the $N$th resonance of the clean ring do not overlap with the frequencies arising from the splitting of the $(N+1)$th resonance. This yields the condition $N\phi \ll 1$.

Finally, we assumed that the radiation is sufficiently weak so that all four resonances predicted by Eq.~\eqref{JNNN} are well separated. For weak disorder $U_n   \ll  D_n$ the latter requrement is satisfied if $\alpha \ll  \phi$.

\begin{figure}[t]
\centerline{\includegraphics[width=0.30\textwidth]{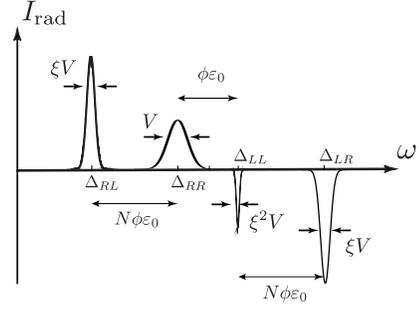} }
\caption{Disorder-induced splitting   of  $N-$th resonance  into four peaks. Amplitudes of $RL$ and $LR$ peaks are enhanced by a factor $ N\approx n_F.$}
\label{resonansy}
\end{figure}

The structure of resonances, which follows from Eq.~\eqref{JNNN}, is shown in Fig.~\eqref{resonansy} assuming the limit of weak disorder $|\xi_n| \ll 1$ (or $\lambda_n \simeq 1$). In this limit one can neglect the disorder-induced level repulsion hence the resonance frequencies are set by
\beml
\beq
&&\Delta_{RR}=\Delta_N, \qquad  \Delta_{LL}= \Delta_N +2\varepsilon_0\phi, \\
&&\Delta_{LR}=\Delta_N  +2 (N+1)\varepsilon_0  \phi,\\
&& \Delta_{RL}= \Delta_N-2N \varepsilon_0 \phi.
\eq
\eml
The width and height of each of the resonances are determined by the corresponding matrix element $V_{ab}$. The current direction at resonance is also different.  Remarkably, the amplitudes of $LR$ and $RL$ resonances are enhanced by a factor $N \approx n_F$ as compared to  those of $RR$ and $LL$ resonances provided weak disorder regime $|\xi_n|  \ll 1/N$.  In this regime the $LR$ and $LL$  transitions correspond to antiresonances, i.e,the direction of current is opposite to those at $RL$ and  $RR$ resonances. Thus, weak short-range disorder does not suppress or smoothen the $RR$ resonant peak in current but leads instead to the appearance of two sharp and intense resonances of opposite chirality corresponding to $LR$ and $RL$ transitions.

For stronger but still sufficiently weak disorder, such that  $1/N \ll  |\xi_n|  \ll 1$, the amplitudes and signs of the $LR$ and $RL$ resonant peaks do not change, while the $RR$ and $LL$ resonances are strongly enhanced. The corresponding amplitudes are of opposite sign and proportional to $\pm N( |\xi_N|^2- |\xi_{N+1}|^2)$.  The absolute sign of the factor $ |\xi_N|^2- |\xi_{N+1}|^2$ depends, however, on disorder realization and cannot be predicted.  Thus, resonant optical excitations in the ring can be used to probe mesoscopic fluctuations of disorder. More specifically, the sign of $RR$ resonance might change for different disorder realizations.

Finally, we notice that the amplitudes of all $4$ resonant peaks decreases provided disorder becomes so strong that $U_n\gg D_n$ and, consequently, $\lambda_{n} \ll 1$.

\begin{figure}[t]
\centerline{\includegraphics[width=0.30\textwidth]{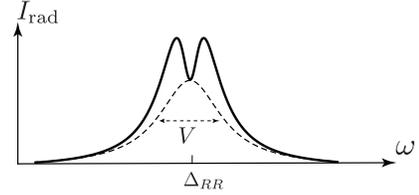} }
\caption{Structure of $N-$th resonance  for $N\phi \ll \alpha$.}
\label{resonansy-phi0}
\end{figure}

With increasing radiation intensity individual resonances start to overlap. For $\alpha \gg \phi$  only $RR$ and $LL$ resonances overlap while $LR$ and $RL$ resonances remain well separated. For $\alpha \gg  N \phi$ all $4$ peaks overlap and form a wide resonance which in the first approximation is described by Eq.~\eqref{JNN}. The effect of weak disorder remains small since the levels $n$ and $-n$ are no longer degenerate even at $\phi=0$ due to radiation-induced level  repulsion. In this case, the effect of  disorder can be taken into account by the standard perturbative analysis using Eqs.~\eqref{chiN} and \eqref{chiN+1} as zero approximation. Calculating the perturbative corrections up to second order with respect to disorder potential we obtain
\beq
J_N&=& \frac{I_0}{2} \frac{\alpha^2}{\sqrt{\alpha^2+\delta_N^2}(|\delta_N|+ \sqrt{\alpha^2+\delta_N^2})} \nonumber \\
&\times& \left [ 1+ \frac{8N (|U_N|^2 - |U_{N+1}|^2)}{\varepsilon_0^2 (|\delta_N|+ \sqrt{\alpha^2+\delta_N^2})^2}+\ldots\right].
\label{JNNU}
\eq
This result is valid for $\alpha \gg \textrm{max}\,\{N\phi, \sqrt{N} |U_n|/\varepsilon_0\}$. Interestingly, the sign of disorder-induced correction to current also depends on particular realization of random potential. For the case $|U_N| < |U_{N+1}|$ the resulting current is plotted schematically in the Fig.~\ref{resonansy-phi0} as a function of frequency.

%%%%%%%%%%%%%%%%
%%%%% Fig. 8 resonansy
%%%%%%%%%%%%%%%%
%%%%%%%%%%%%%%%%

%%%%%%%%%%%%%%%%
%%%%% Fig. 9 resonansy-phi0
%%%%%%%%%%%%%%%%
%%%%%%%%%%%%%%%%

The case of nearly half-integer flux piercing the ring may be considered in a similar fashion. Assuming that $\phi \approx 1/2$ we find that the energy levels are also double degenerate, but the disorder potential $U(\varphi)$ mixes the level $N$, which has a positive chirality, with the level $-(N+1)$, which has a negative chirality. Corresponding matrix elements differ slightly from those given by Eq.~\eqref{UNN},
\be
U_N=\frac{1}{2\pi}\int_0^{2\pi} U(\varphi) e^{-2i(N+1)\varphi}d\varphi,
\ee
which does not make, however, a difference for the structure of resonances. Still, the width, height and position of resonant peaks change accordingly.

\subsection{Long-range disorder}

In this subsection we consider the effect of static long-range disorder $U(\varphi)$ that does not lead to scattering between left- and right- moving electron states. Consequently it is still convenient to analyze the effect in the rotating reference frame. The corresponding electron wave function $\chi(\theta)$ yields the Schr{\"o}dinger equation (\ref{Shr-stat}) which is equivalent to those for a quantum physical pendulum. In contrast to the laboratory frame,  the corresponding  potential in the rotating frame is no longer static but oscillates with the frequency $\omega$,
\be
U(\varphi)=U(\theta+\omega t)= \sum \limits_{ n=-\infty}^{\infty}\!\!\!  U_n \,e^{in(\theta+\omega t)}.
\label{Ut}
\ee
Since the potential is smooth, i.e.,it does not change essentially on the scales of the order of the electron wavelength, the quasiclassical analysis is justifiable. The problem is, therefore, reduced to that of a classical physical pendulum subject to a fast-oscillating potential. Such a model is often considered in textbooks as the simplest example of a system that shows chaotic behavior.\cite{Zaslavsky}

It is well known that the effect of the fast-oscillating potential is negligibly small everywhere except for a narrow strip in the phase space that "dresses" the separatrix (a curve which separates oscillating and rotating pendulum states). Such a strip is called the chaotic layer.  Within the chaotic layer the physical pendulum jumps randomly between dynamically localized and delocalized trajectories (in the phase space) thus showing a chaotic behavior. The width of the chaotic layer can be estimated as\cite{Zaslavsky}
\be
\Gamma_{\rm ch} \propto  |U_1|\exp{\left(-\frac{\pi\omega}{\Omega}\right)},
\label{Gamma-ch}
\ee
where $U_1$ is the amplitude of the first  Fourier harmonic of the oscillating potential [see Eq.~\eqref{Ut}] and  $\Omega$ is the characteristic energy scale corresponding to pendulum frequency at the point of equilibrium.

%%%%%%%%%%%%%%%%
%%%%% Fig. 10 current_chaotic
%%%%%%%%%%%%%%%%
\begin{figure}[t]
\centerline{\includegraphics[width=0.40\textwidth]{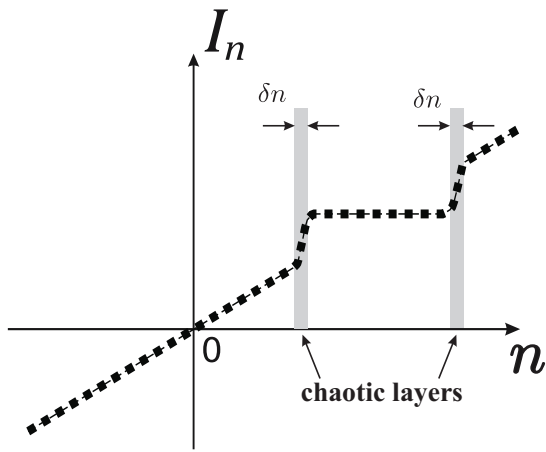} }
\caption{Levels captured in the chaotic layer (shown in gray) are randomly populated and depopulated  leading to current fluctuations.}
\label{current_chaotic}
\end{figure}
%%%%%%%%%%%%%%%%

The analogy between a quantum disordered ring subjected to circularly polarized light and the physical pendulum in oscillating potential suggests that the long-range disorder may play a role only in the limit of strong coupling to electromagnetic field, i.e., for $\alpha \gg 1$.  In this case the number of levels perturbed by the radiation-induced potential $W$ is proportional to $\sqrt \alpha \gg 1$. This justifies the quasiclassical approach which suggests that $\Omega=\sqrt{\alpha} \varepsilon_0$ is the characteristic energy scale which enters Eq.~\eqref{Gamma-ch}.

The chaotic layer separates the regions with localized and delocalized states.  For strong coupling the chaotic layer is confined to an energy interval around $\varepsilon =0$ of the width $\delta \varepsilon \simeq \Gamma_{\rm ch}$  (see   Fig.~\ref{pendulum}). The energy levels captured by the chaotic region correspond to the values of $n$ such that $|n-n_\omega| \simeq \sqrt \alpha$ as shown in Fig.~\ref{current_chaotic}.  The number of levels within the chaotic layer can be estimated as
\be
\delta n \simeq  \rho_{\rm ch}   \Gamma_{\rm ch},
\label{del-n}
\ee
where $\rho_{\rm ch}$ is a density of electron states within the chaotic layer. Simple quasiclassical analysis of Eqs.~\eqref{Shr-stat} and  \eqref{bound} shows that   $\rho(\varepsilon) \simeq \ln(\alpha/\varepsilon)/\varepsilon_0 \sqrt{\alpha}$ in a vicinity of the separatrix, so that
$\rho_{\rm ch}$ can be estimated as  $\rho_{\rm ch} \simeq \ln(\alpha\varepsilon_0/\Gamma_{\rm ch})/\varepsilon_0 \sqrt{\alpha}.$

The chaotic behavior leads to  fluctuations of dc current due to random jumps within the chaotic layer. The amplitude of such fluctuations is given by
\be
\label{DI}
\Delta I \simeq \frac{\partial I_n}{\partial n} \delta n.
\ee
In order to estimate the derivative $\p I_n/\p n$ we take advantage of the perturbative result given by Eq.~\eqref{dIn} which is taken at the boundary of its applicability range, i.e.\,for $|n-n_\omega| = \sqrt \alpha$. In this way we find $\p I_n/\p n \approx I_0$. Assuming that $\Gamma_{\rm ch} \ll \alpha   \varepsilon_0$ we obtain from Eqs.~(\ref{Gamma-ch}-\ref{DI}) that
\be
\Delta I \simeq I_0 \frac{\Gamma_{\rm ch}}{\varepsilon_0}\frac{\ln{\left(\alpha\varepsilon_0/\Gamma_{\rm ch}\right)}}{\sqrt{\alpha}}.
\ee

The fluctuations of current due to chaotic dynamics reveal themselves as a burst noise, i.e.\,a random telegraph signal.  An electron captured within the chaotic layer makes a random hop during a time $ \rho_{\rm ch}$ which is a characteristic time required for an electron to travel along a typical trajectory within the chaotic layer. Sudden jumps in the current of a magnitude $\Delta I$ correspond to a hopping event for any of $\delta n$ available electrons.  Thus, the characteristic rate of current jumps can be estimated as $1/\tau_{\rm ch} \approx  \delta n / \rho_{\rm ch}.$  Hence $\tau_{\rm ch} \simeq {1}/{\Gamma_{\rm ch}}$ as can be seen from Eq.~\eqref{del-n}.

\section{Ballistic multi-channel ring} \label{many}

The generalization of our results to the case of ballistic multi-channel rings is rather straightforward. In the absence of radiation the energy levels in a multi-channel ring are  given by
 \be
 E_{nm} ^{(0)}= \varepsilon_0 \frac{n^2}{2} +\varepsilon_\perp ^m,
 \ee
where the additional index $m$ numerates the subbands (channels) due to transverse quantization. The specific expression for subband energies $\varepsilon_\perp ^m$ depends on the confining electrostatic potential in the transverse direction.

The contribution of the $m$th subband to optically induced {\it dc } current can be easily found from equations obtained in the previous sections by replacing the Fermi energy with the energy
\be
 E_F^m= E_F -  \varepsilon_\perp ^m,
\ee
and by redefining the parameters such that
\beml
\label{subst1}
\beq
n_F  &\to& n_F^m=\sqrt{2E_F^m/\varepsilon_0},   \label{nFm}\\
\Delta_F  &\to& \Delta_F^m=\varepsilon_0 n_F^m, \label{DeltaFm}\\
\delta \omega &\to& \delta\omega^m=T/n_F^m. \label{domegam}
\eq
\eml
For weak coupling $\alpha \ll 1$ the position of resonances would correspond to the energies  $E_{n+1,m} ^{(0)}- E_{nm} ^{(0)}$. For each channel $m$ the transverse energy drops out from this expression hence the parameters $\delta_n$ entering Eq.~\eqref{I-rad-final} are still given by Eq.~\eqref{delta-small-n}. The existence of several conducting channels would affect, however, the envelope function which is given by Eq.~\eqref{env} for a single-channel ring. This result is, however, readily generalized with the help of the substitutions \eqref{subst1}. Performing the summation over all subbands we obtain the current
 \be
I_{\rm env} (\omega) \approx \frac{I_0}{8T} \sum \limits_{m=0}^{m_F}
\frac{\Delta_F^m}{\cosh^2[(\omega-\Delta_F^m)/2\delta\omega^m]},
\label{envm}
 \ee
 where $m_F$ is found from the equation $\varepsilon_\perp^{m_F}=E_F$.

The resonant peaks in the envelope function $I_{\rm env} (\omega)$, which correspond to the $m$th and $(m+1)$th  subbands, are well separated as far as  $\Delta_F^{m+1}-\Delta_F^{m} \gg \delta \omega^m$. The latter inequality is equivalent to the condition
\be
 \frac{\p \varepsilon_\perp^m}{\p m} \gg T,
 \label{ineqm}
 \ee
which can be obtained with the help of Eqs.~\eqref{subst1} using that $\Delta_F^{m+1}-\Delta_F^{m}\approx \p\Delta_F^{m}/\p m$. Thus, the subbands provide well separated resonant contributions to the current provided the temperature is small compared to the intersubband spacing.

For higher temperatures inequality \eqref{ineqm} is violated and the resonant peaks in the envelop function $I_{\rm env} (\omega)$ start to overlap.  Finally, at very large temperatures, such that $T \gg   {\p \varepsilon_\perp^m}/{\p m}$ for any $m < m_F$, all resonances merge in a single wide peak which can be described by Eq.~\eqref{envm},
 where the summation over $m$ is replaced with integration,
\be
I_{\rm env} (\omega) \approx \frac{I_0}{8T} \int \limits_{\varepsilon_\perp^m<E_F}  dm \;\frac{\Delta_F^m}{\cosh^2[(\omega-\Delta_F^m )/2\delta\omega^m]}.
\label{envm1}
 \ee
The integral is readily estimated for the simplest model of the ring of a finite width $a$ assuming that the effect of the confining potential is properly accounted by using periodic boundary conditions in the transversal direction. For such a model we obtain
\be
\varepsilon_\perp ^m=\varepsilon_\perp  \frac{m^2}{2},
\ee
where $\varepsilon_\perp=(2\pi \hbar)^2/Ma^2$. We also find
\beml
\beq
E_F^m &=&E_F(1-m^2/m_F^2),\\
\Delta_F^m &=&\Delta_F\sqrt{1-m^2/m_F^2},\\
\delta \omega^m &=& \delta \omega/\sqrt{1-m^2/m_F^2},
\eq
\eml
where
\be
m_F=\sqrt{{2E_F}/{\varepsilon_\perp}}.
\ee
Using that $\Delta_F \gg \delta \omega$ (this inequality is fulfilled for $E_F \gg T$) one can find the asymptotic behavior of the integral in Eq.~\eqref{envm1} as
\be
 \label{env2}
I_{\rm env} (\omega)\approx \frac{I_0 \sqrt{T/ \varepsilon_\perp}}{4n^*}
\begin{cases}
\sqrt{ 2\pi} e^{\frac{\Delta_F-\omega}{\delta \omega}}, & \omega-\Delta_F \gg \delta \omega,  \\
C,& |\omega-\Delta_F| \ll \delta \omega, \\
\frac{2\omega\sqrt{\delta\omega/\Delta_F}}{ \sqrt{\Delta_F^2 -\omega^2}}, &
\Delta_F - \omega \gg \delta \omega,
\end{cases}
\ee
where $C=\int_0^\infty dx/\cosh^2(x^2) \approx 0.95.$

The result of Eq.~(\ref{env2}) predicts exponential decay of current for large frequencies,  $\omega-\Delta_F \gg \delta \omega$. One can, however, see that such behavior is limited by $\omega< \Delta_F + \delta \omega \ln(E_F/T)$.  For larger values of $\omega,$  $I_{\rm env}$ decays in a slower power-law way.

Comparing Eq.~\eqref{env2} with Eq.~\eqref{env} we find that at high temperatures the optical response in the multichannel ring is enhanced by a factor $\sqrt{T/{\varepsilon_\perp}}$.

\section{Discussion and conclusion}
\label{discussion}

In this section we estimate the value of the current $I_{\rm rad}$ induced in a semiconducting nanoring by circularly polarized light. We also discuss related problems which are to be addressed in future, and summarize the results obtained.

The only parameter of the theory which depends on specific material properties is the electron mass $M$. To make estimates we take the standard value of the effective mass for GaAs, $M\approx 0.07 m_0$, where $m_0$ is the free electron mass. We also take the radius of the ring to be $R=100$\,nm. For a single-channel ring we, then, obtain using these parameters that $\varepsilon_0\approx 10^{-4}$\,eV  and $I_0 \approx 3.7\times  10^{-9}$\,A. For a wide range of  Fermi energies $E_F = 0.02 - 2$\,eV,  we find $n_F \approx 20 - 200$ and $\Delta_F\approx (0.2 - 2) \times 10 ^{-2}$\,eV. We see that inequality Eq.~\eqref{TDF} can be easily satisfied for not too large temperatures. The corresponding  resonance frequency turns out to be in the THz range, $f=\omega/2\pi= \Delta_F/2\pi  \approx 0.5 - 5$\,THz. The coupling to electromagnetic field becomes stronger for sufficiently low fields such that $\alpha =1$ for $\mathcal E_0 \approx 10$\,V/cm. The maximal value of the current in the weak coupling regime  (this value is reached in $LR$ and $RL$ disorder-induced resonances)  is estimated as $I_0n_F \approx (0.8- 8) \times 10^{-7}$\,A. In the  very strong coupling regime, $\alpha\gg n_F^2$, the current increases to reach a maximal value $I_{\rm max} \simeq I_0 n_F^2 \approx  (1.5 - 150) \times 10^{-6}$ A.  The conditions for the very  strong coupling regime are satisfied only for sufficiently  strong fields: $\mathcal E_0 \gtrsim (4 - 400) \times 10^3$\,V/cm.

Let us now estimate the magnetic field induced by the current $I_{\rm rad}$.  For a single ring, the field in the center of the ring is given by a simple formula
\be
B=\frac{2\pi I_{\rm rad}}{c R},
\ee
where $c$ is the speed of light in the vacuum.  In the weak-coupling regime, $I_{\rm rad} \simeq I_0n_F$, we estimate  $B\approx (0.5-5) \times 10^{-6}$\,T.  In the strong-coupling regime the maximal field $B_{\rm max} \approx (0.1-1) \times 10^{-3}$\,T is reached for $I_{\rm rad} \simeq  I_{\rm max}$. This field can increase further in a multi-channel ring and/or by using three-dimensional arrays of rings. Another way to increase the effective magnetic field generated by the ring is to make the Fermi energy and, consequently, the parameter $n_F$ larger.

In our analysis we focused on the dependence of the $I_{\rm rad}$ on the frequency $\omega$ of incoming radiation and found that circular current might show sharp resonances. Importantly, the dependence of  $I_{\rm rad}$ on the magnetic flux $\phi$ also reveals sharp peaks for a given $\omega$ provided interaction constant $\alpha$ is sufficiently small. Indeed, the positions of resonances, which are shown in the upper panel of Fig.~\ref{current-both}, depend on magnetic flux. Increasing magnetic flux by the flux quantum, $\phi\to\phi+1$, is equivalent to the substitution $\delta_n\to\delta_{n-1}$ in Eq.~\eqref{I-rad-final} that describes the { dc} photoresponse for the case of adiabatic radiation switching.  (The same remains true for the ring coupled to a thermal bath provided weak coupling to electromagnetic radiation.) Thus, the current $I_{\rm rad}$ is a periodic function of $\phi$ with the period $1$ as expected.  It is evident from the consideration above that there exists a single sharp peak in $\phi$ dependence of $I_{\rm rad}$ in the interval $0<\phi<1$. The ratio of the maximal value of the current, $I_{\rm rad}^{\rm max}$ in this interval to the flux-averaged current $\langle I_{\rm rad}\rangle_{\phi}$ is as large as $1/\alpha$ and $1/\sqrt{\alpha^2+\Gamma^2/2}$ for the case of adiabatic radiation switching and thermal bath coupling, respectively. This ratio also gives an estimate for the number of harmonics that are effectively contributing to the Fourier expansion of the radiation-induced current $I_{\rm rad}=\sum_m I_m \exp(2\pi i m \phi ).$  The bigger the ratio the larger the number of relevant harmonics  with a large amplitude that can be observed in experiment.

The dependence of the circular current on $\phi$  is essentially different in the strong-coupling regime, $\alpha \gg 1$.   In this  case,  all harmonics $I_m$ for $m\neq 1$ are small compared to $I_0$.  In particular, $I_m/I_0 \propto \exp(-2\pi^2 |m| n^*)$  or $I_m/I_0 \propto \exp(-2\pi |m| \Gamma)$  for  adiabatic radiation switching (assuming $1\ll n^* \ll \sqrt \alpha  $) and  for the case of thermal bath coupling (assuming $\Gamma=\Gamma_\varphi,$ and $1<\alpha<\Gamma^2$), respectively.\cite{comm2}    Hence, in the case of strong coupling to the radiation the response  is given by a large flux-independent  quasiclassical contribution and a small  quantum correction oscillating with $\phi.$ The latter is dominated by the contribution of  harmonics with $m=\pm1.$

In the presented analysis we ignored the effects of the electron-electron interactions that may not be negligible. Even thought the detailed study of interaction-induced effects is a complex task that falls outside the scope of the current paper some qualitative predictions can be already made.

One may expect that in a single-channel ring at sufficiently low temperatures the main effect of electron-electron interactions is to renormalize the value of the coupling strength $\alpha$. Such renormalization will likely result in the suppression of $\alpha$ by a factor $(T/E_F)^{g^2}$, which is characteristic for the Luttinger liquid behavior, where $g$ is a dimensionless interaction constant. This effect can be taken into account by replacing the coupling constant $\alpha$ with its renormalized value that would not change essentially the predictions of our analysis.

If interactions are sufficiently strong, less trivial effects, which are related to the charge  quantization in a finite geometry, may show up.  As was first demonstrated in Ref.~\onlinecite{Dmitriev10}  the electron-electron interactions in a 1D ring give rise to an effective contribution to magnetic flux that is proportional to both the interaction constant $g$ and the imbalance $N_R-N_L$, where $N_R (N_L)$ is the total number of right-(left-) moving electrons in an ideal ring. As a result sufficiently strong interactions in a clean system would lead to further splitting of the four resonances described in Sec.~\ref{disorder}.

A completely different but sizable effect of interactions is expected in multi channel rings for sufficiently high temperatures such that the electron-electron collisions dominate. This case is generally referred to as the hydrodynamic regime. In this regime elementary excitations in the ring are dominated by plasmons. The corresponding plasmonic resonance in the dc current has a width which is much smaller than the resonance width in the ballistic noninteracting ring studied above. The decrease of the linewidth is due to the motional line narrowing caused by intense electron-electron collisions.  Such and other interaction-related phenomena will be studied elsewhere.

To conclude, we developed a theory of the inverse resonant Faraday effect in quantum rings. We demonstrated that a circularly polarized radiation with the frequency $\omega$ induces a dissipationless  dc current  $I_{\rm rad}$ in a quantum ring pierced by magnetic flux $\phi$. The current yields the symmetry, Eq.~\eqref{Irad_phi_omega}, so that  the direction of the optically-induced current  is sensitive  to helicity of the incoming radiation.

For the case of weak coupling the current $I_{\rm rad}(\omega,\phi)$ reveals sharp resonances as a function of $\omega$ for a given flux $\phi$. These resonances can also be observed by changing the flux for a fixed frequency of light $\omega$.  Analytical expressions for the radiation-induced current are obtained for two different cases: (i) an isolated ring under the assumption of adiabatic switching of light intensity, and (ii) a quantum ring weakly coupled to the thermal bath.

The nonresonant current is found to be proportional to the squared amplitude of light $I_{\rm rad} \propto \mathcal E_0^2$ in agreement with the conventional theory of the nonresonant inverse Faraday effect.  The current is, however, strongly enhanced at a resonance so that its maximal value does not depend on the intensity of light (in the regime when dissipation is negligible).

For the case of strong coupling multiple resonances in $I_{\rm rad}(\omega,\phi)$  merge into a wide peak with a width determined by the spectral curvature.  The amplitude of the peak increases with $\mathcal E_0$ as $\mathcal E_0^{3/2}$ for small light intensity. It is proportional to $\mathcal E_0$ for moderate intensities and finally saturates in the limit of hight intensity of light. The saturated value of the current scales as $n_F^2$ with the total number of electrons in the ring  $n_F$.

Weak disorder is shown to affect the dependence of $I_{\rm rad}$ on frequency in a highly nontrivial way. In contrast to naive expectations a weak short-range disorder does not suppress resonances but leads instead to the appearance of additional resonant peaks of different polarity.  These sharp resonant features are suppressed only by relatively strong disorder potential. Thus, we find that the inverse Faraday effect is generally very sensitive to the quality of the ring.

 The long-range disorder does not affect the picture for the case of weak coupling to light while it becomes essential for the case of strong coupling. The main effect of long-range disorder is to induce a chaotic behavior of the system in the vicinity of the separatrix that divides the phase space into the regions with dynamically localized and delocalized states. The radiation-induced current $I_{\rm rad}$  is shown to fluctuate due to random electron hopping within the narrow chaotic layer in the phase space of the system, which "dresses" the separatrix.  Such fluctuations lead to the burst noise in the optical dc response and power dissipation.

Finally, we generalize some of the results obtained for the case of a multichannel ring.  We demonstrate that at low temperatures the response originating in different propagation channels is well separated in frequency so that the spectrum of different subbands can be resolved in experiment.  At higher temperatures the resonances overlap but the overall response is enhanced by a factor $\sqrt{T/\varepsilon_\perp}$ as compared to the case of a single-channel ring.

\section{Acknowledgements}
We thank I.\,V.~Gornyi and M.\,E.~Portnoi for useful discussions.  The work was supported by the Dutch Science Foundation NWO/FOM 13PR3118, by the EU Network FP7-PEOPLE-2013-IRSES Grant No 612624 ``InterNoM'', by Russian Foundation For Basic Research (14-02-00198),  by Programs of RAS,  and by the U.S. Army Research Laboratory through the Collaborative Research Alliance for Multi-Scale Modelling of Electronic Materials.


\begin{thebibliography}{99}

\bibitem{Dai15}%Extracting entangled qubits from Majorana fermions in quantum dot chains through the measurement of parity
L.~Dai, W.~Kuo and Ming-Chiang Chung,
Scientific Reports \textbf{5}, 11188 (2015).

\bibitem{Liu15}%Direct measurement on the geometric phase of a double quantum dot qubit via quantum point contact device
B.~Liu, F.-Y.~Zhang, J.~Song and H.-S.~Song,
Scientific Reports \textbf{5}, 11726 (2015).

\bibitem{Kostarelos14}%Exploring the Interface of Graphene and Biology
K.~Kostarelos and K.\,S.~Novoselov,
Science \textbf{344}, 261 (2014).

\bibitem{Jayich09}%Persistent Currents in Normal Metal Rings
A.\,C.~Bleszynski-Jayich, W.\,E.~Shanks, B.~Peaudecerf, E.~Ginossar, F.~von Oppen, L.~Glazman, and J.\,G.\,E.~Harris,
Science \textbf{326}, 272 (2009).

\bibitem{Birge09}%Sensing a Small But Persistent Current
N.\,O.~Birge, Science \textbf{326}, 244 (2009).

\bibitem{Srivastava15}%Optically active quantum dots in monolayer WSe2
A.~Srivastava,	M.~Sidler,	A.\,V.~Allain, D.\,S.~Lembke, A.~Kis	and A.~Imamoglu,
Nature Nanotechnology \textbf{10}, 491 (2015).

\bibitem{Fang15}%Nanoplasmonic waveguides: towards applications in integrated nanophotonic circuits
Y.~Fang and M.~Sun, Light: Science \& Applications \textbf{4}, e294 (2015).

\bibitem{Beaulac09}%Light-Induced Spontaneous Magnetization in Doped Colloidal Quantum Dots
R.~Beaulac, L.~Schneider, P.\,I.~Archer, G.~Bacher and D.\,R.~Gamelin,
Science \textbf{325}, 973 (2009).

\bibitem{Cave09}%Inducing Chirality with Circularly Polarized Light
R.\,J.~Cave, Science \textbf{323}, 1435 (2009).

\bibitem{Borunda08}%Aharonov-Casher and spin Hall effects in two-dimensional mesoscopic ring structures with strong spin-orbit interaction
M.\,F.~Borunda, X.~Liu, A.\,A.~Kovalev, X.-J.~Liu, T.~Jungwirth, J.~Sinova, Phys.\ Rev.\ B \textbf{78}, 245315 (2008).

\bibitem{Oudenaarden98}%Magneto-electric Aharonov?Bohm effect in metal rings
A.~van Oudenaarden, M.\,H.~Devoret, Yu.\,V.~Nazarov, J.\,E.~Mooij, Nature \textbf{391}, 768 (1998)

\bibitem{Webb85}%Observation of h/e Aharonov?Bohm Oscillations in Normal-Metal Rings
R.\,A.~Webb, S.~Washburn, C.\,P.~Umbach, R.\,B.~Laibowitz, Phys.\ Rev.\ Lett. \textbf{54}, 2696 (1985).

\bibitem{Schoenenberger99}%Aharonov?Bohm oscillations in carbon nanotubes
C.~Sch\"onenberger, A.~Bachtold, C.~Strunk, J.-P.~Salvetat, J.-M.~Bonard, L.~Forr\'o, T.~Nussbaumer, Nature \textbf{397}, 673 (1999).

\bibitem{Grbic08}%Aharonov-Bohm oscillations in p-type GaAs quantum rings
B.~Grbic, R.~Leturcq, T.~Ihn, K.~Ensslin, D.~Reuter, A.\,D.~Wieck, Physica E \textbf{40}, 1273 (2008).

\bibitem{Titov97}%Log-normal Distribution of Level Curvatures in the Localized Regime: Analytical Verification
M.~Titov, D.~Braun, Y.\,V.~Fyodorov, J.\ Phys. A \textbf{30}, L339 (1997).

\bibitem{Dyakonov93}
M.\,I.~Dyakonov and M.\,S.~Shur, Phys. Rev. Lett. {\bf 71}, 2465 (1993).

\bibitem{Ivchenko2011}
E.\,L.~Ivchenko and S.\,D.~Ganichev, Pisma v ZheTF {\bf 93}, 752 (2011) [JETP Lett. {\bf 93}, 673 (2011)].

\bibitem{Aleiner02}%Quantum Effects in Coulomb Blockade
I.\,L.~Aleiner, P.\,W.~Brouwer and L.\,I.~Glazman, Phys.\ Rep. \textbf{358}, 309 (2002).

\bibitem{Evers08}%Anderson transitions
F.~Evers and A.\,D.~Mirlin, Rev.\ Mod.\ Phys. \textbf{80}, 1355 (2008).

\bibitem{Kouwenhoven98}%Quantum Dots
L.\,P.~Kouwenhoven and C.\,M.~Marcus, Physics World \textbf{11}, 35 (1998).

\bibitem{Pitaevskii61}
L.\,P.~Pitaevskii, Sov.\ Phys.\ JETP \textbf{12}, 1008 (1961).

\bibitem{Ziel65}
J.\,P.~van der Ziel, P.\,S.~Pershan and L.\,D.~Malmstrom, Phys.\ Rev.\ Lett. \textbf{15}, 190 (1965).

\bibitem{Kimel05}
A.\,V.~Kimel, A.~Kirilyuk, P.\,A.~Usachev, R.\,V.~Pisarev, A.\,M.~Balbashov and Th.~Rasing, Nature \textbf{435}, 655 (2005).

\bibitem{Kirilyuk10}
A.~Kirilyuk, A.\,V.~Kimel, and T.~Rasing, Rev.\ Mod.\ Phys. \textbf{82}, 2731 (2010).
\bibitem{Kirilyuk11}%Controlling spins with light
A.~Kirilyuk, A.\,V.~Kimel and Th.~Rasing, Phil.\ Trans.\ R.\ Soc.\ A \textbf{369}, 3631 (2011).

\bibitem{Kibis11}
O.\,V.~Kibis,  Phys.\ Rev.\ Lett.  \textbf{107}, 106802 (2011).

\bibitem{Kibis13} O.\, V.~ Kibis, O.~Kyriienko,   I.\, A.~ Shelykh,  Phys.\ Rev.\ B  \textbf{ 87}, 245437 (2013).

\bibitem{Alexeev13}%Aharonov-Bohm quantum rings in high-Q microcavities
A.\,M.~Alexeev, I.\,A.~Shelykh, M.\,E.~Portnoi, Phys.\ Rev.\ B \textbf{88}, 085429 (2013).

\bibitem{Joibari14}%Light-induced spin polarizations in quantum rings
F.\,K.~Joibari, Ya.\,M.~Blanter, G.\,E.\,W.~Bauer,
Phys.\ Rev.\ B \textbf{90}, 155301 (2014).

\bibitem{Alexeev12}  A.\,M.~Alexeev,  M.\,E.~Portnoi, Phys.\ Rev.\ B \textbf{85}, 245419 (2012).
\bibitem{Kruglyak2005}V.\, V.~ Kruglyak, M.\, E.~ Portnoi,  Technical Physics Letters, \textbf{ 31},  1047  (2005)  [Pis’ma v Zh. Tekh. Fiziki  \textbf{31}, 20  (2005).
\bibitem{Kruglyak2007}V.\,V.~Kruglyak, M.\, E.~Portnoi, R.\,J.~Hicken,  Journal of Nanophotonics, \textbf{ 1}, 013502 (2007).
\bibitem{Polianski2009} M.\,L.~Polianski,  Phys.\ Rev.\ B \textbf{80}, 241301(R) (2009).

\bibitem{Imry}
Y.~Imry,  {\it Introduction to mesoscopic physics},  Oxford, Oxford University Press (2002).

\bibitem{Altshuler91} B.\, L.~Altshuler,  Y.~Gefen, Y.~Imry,    Phys.\ Rev.\ Lett.  \textbf{66}, 88 (1991).

\bibitem{Dmitriev10}
A.\,P.~Dmitriev, I.\,V.~Gornyi, V.\,Yu.~Kachorovskii  and D.\,G.~Polyakov,
Phys.\ Rev.\ Lett. \textbf{105}, 036402 (2010).

\bibitem{Shmakov12}P. \,M.~ Shmakov, A.\, P.~Dmitriev, V.\,Yu.~ Kachorovskii,   Phys. Rev. B  \textbf{85}, 075422  (2012);
Phys.\ Rev.\ B \textbf{87}, 235417 (2013).

\bibitem{Dmitriev14}
A.\,P.~Dmitriev, I.\,V.~Gornyi, V.\,Yu.~Kachorovskii, D.\,G.~Polyakov, P.\,M.~Shmakov,  JETP Letters, \textbf{100},  946  (2014).

\bibitem{Levy90}
L.\,P.~Levy, G.~Dolan, J.~Dunsmuir, and H.~Bouchiat,  Phys.\ Rev.\ Lett. \textbf{64}, 2074 (1990).

\bibitem{Reulet95}
B.~Reulet, M.~Ramin, H.~Bouchiat, D.~Mailly, Phys.\ Rev.\ Lett. \textbf{75}, 124 (1995).

\bibitem{Deblock02}
R.~Deblock, R.~Bel, B.~Reulet, H.~Bouchiat, and D.~Mailly,  Phys.\ Rev.\ Lett. \textbf{89}, 206803 (2002).

\bibitem{Kleemans07}
N.\,A.\,J.\,M.~Kleemans, I.\,M.\,A.~Bominaar-Silkens, V.\,M.\,~Fomin, V.\,N.\,~Gladilin, D.\,~Granados, A.\,G.\,~Taboada, J.\,M.\,~Garcia, P.\,~Offermans, U.\,~Zeitler, P.\,C.\,M.\,~Christianen, J.\,C.\,~Maan, J.\,T.\,~Devreese, and P.\,M.\,~Koenraad, Phys.\ Rev.\ Lett. \textbf{99}, 146808 (2007).

\bibitem{linear} It worth noting that for linearly polarised radiation the situation is different because in this case the averaged optically-induced current is exponentially small and is sensitive to mesoscopic fluctuations in full analogy with persistent current.

\bibitem{Zaslavsky}
G.\,M.~Zaslavsky,  "{\it The physics of chaos in Hamiltonian systems.}",   Imperial College, London (2007).
\bibitem{comm2} More detailed calculation of $I_m$ in the different cases  requires  analysis of analytical properties of $J_n$ considered as a function of $n.$ Such analysis is out of scope of the current paper and will be presented elsewhere.
\end{thebibliography}
\end{document}